\documentclass[fleqn,usenatbib]{mnras}

\usepackage{newtxtext,newtxmath}

\usepackage[T1]{fontenc}

\DeclareRobustCommand{\VAN}[3]{#2}
\let\VANthebibliography\thebibliography
\def\thebibliography{\DeclareRobustCommand{\VAN}[3]{##3}\VANthebibliography}

\usepackage{graphicx}	
\usepackage{amsmath}	
\usepackage{amssymb}	

\title[Stability and gravitational collapse of neutron stars with realistic equations of state]{Stability and gravitational collapse of neutron stars with realistic equations of state}

\author[J. Z. Pretel and M. A. da Silva]{
J. M. Z. Pretel,$^{1}$\thanks{E-mail: juanzarate@if.ufrj.br}
and M. F. A. da Silva,$^{2}$
\\
$^{1}$Instituto de F\'\i sica, Universidade Federal do Rio de Janeiro, CEP 21941-972, Rio de Janeiro, RJ, Brazil\\
$^{2}$Departamento de F\'isica Te\'orica, Universidade do Estado do Rio de Janeiro, CEP 20550-013, Rio de Janeiro, RJ, Brazil
}

\date{Accepted XXX. Received YYY; in original form ZZZ}

\pubyear{2020}

\begin{document}
\label{firstpage}
\pagerange{\pageref{firstpage}--\pageref{lastpage}}
\maketitle

\begin{abstract}
We discuss the stability and construct dynamical configurations describing the gravitational collapse of unstable neutron stars with realistic equations of state compatible with the recent LIGO-Virgo constraints. Unlike other works that consider the collapse of a stellar configuration without a priori knowledge if it is stable or unstable, we first perform a complete analysis on stellar stability for such equations of state. Negative values of the squared frequency of the fundamental mode indicate us radial instability with respect to the collapse of the unstable star to a black hole. We find numerical solutions corresponding to the temporal and radial behavior during the evolution of the collapse for certain relevant physical quantities such as mass, luminosity, energy density, pressure, heat flow, temperature and quantities that describe bulk viscous processes. Our results show that the equation of state undergoes abrupt changes close to the moment of event horizon formation as a consequence of dissipative effects. During the collapse process all energy conditions are respected, which implies that our model is physically acceptable. 
\end{abstract}




\section{Introduction}

A neutron star is one of the densest forms of matter in the observable Universe, which is created as a result of gravitational collapse of the central core of a massive star ($> 8 M_\odot$) at the end of his life, followed by a stellar phenomenon known as supernova explosion \citep{LattimerPrakash}. These stars can be detected under a variety of circumstances, usually from optical and X-ray observations \citep{Lattimer}. The internal structure of a neutron star depends on the nuclear equation of state (EoS). Nonetheless, we do not have a perfect theory of cold and ultradense matter, and from the experimental point of view such extreme conditions are not accessible to current laboratory experiments \citep{Ozel}. In the literature there are dozens of different EoSs that are assumed to describe neutron stars \citep{OzelFreire}, so the current strategy is to combine available observations to constrain the underlying EoS of dense matter. Recently the gravitational waves observation from neutron star binary mergers has provided interesting information on the internal composition of these objects, and it is allowing to restrict candidate EoSs for neutron-star matter \citep{Abbott, Abbott1, Schmidt}.

Neutron stars are so compact that it is essential to use general relativity (GR) in order to study their internal structures. In GR it is very well known that stellar configurations in hydrostatic equilibrium correspond to solutions of the Tolman-Oppenheimer-Volkov (TOV) equations. Nevertheless, this equilibrium does not guarantee stability with respect to a compression or decompression caused by a radial perturbation. In fact, there are two conditions for stellar stability \citep{Glendenning}: The necessary condition (also known as the $M(\rho_c)$ method) indicates the first maximum on the $M(\rho_c)$ curve, corresponding to a critical central density from which the stars pass from stability to instability. In other words, hidrostatically stable configurations belong to the region with positive slope (this is, $d M/d\rho_c >0$) on the mass-central density curve. Furthermore, a sufficient condition for stability is to calculate the frequencies of the normal modes of relativistic radial pulsations. If any of these squared frequencies is negative for a particular star, the frequency is purely imaginary and, therefore, the perturbations grow or decrease exponentially with time, i.e. the star is unstable \citep{BardeenTM}. Otherwise, for positive squared frequencies we have stable radial oscillations. If the two conditions for stability give compatible results, then when the central density assumes its critical value, the frequency of the fundamental mode must vanish.

For neutron stars with realistic EoSs, there are two regions in the mass-central density relation that are unstable. On the low densities side there is a instability point (which limits the minimum mass of a neutron star) where the star can become unstable with respect to an explosion, and on the  higher central densities side (where the total gravitational mass $M$ is maximum) the star become unstable with respect to implosion/collapse to a black hole \citep{KokkotasRuoff}. Here we are interested in studying the second case, where the stellar models become unstable with respect to gravitational collapse.

In that respect, some important contributions have been made in the past. On the one hand, the numerical study of radial oscillation modes of neutron stars in GR was carried out by \cite{KokkotasRuoff}, where they present an extensive list of frequencies for various zero temperature equations of state in order to give information about the stability of such stars. On the other hand, the gravitational collapse of neutron stars considered as initial configurations was studied by the authors in \citep{OliveiraPacheSant, MartinezPavon, Martinez, Pretel}. The latter worried about the dynamics of the collapse and transport processes, however, they did not perform an a priori study on the stellar stability of the initial configuration. The EoS plays a fundamental role in determining the internal structure of neutron stars and, consequently, in imposing stability limits. Therefore, it is important first to conduct a study on the stability of the initial static Schwarzschild configurations.

The problem of gravitational collapse was originally approached by  \cite{Oppenheimer}. They investigated the gravitational collapse of a spherically symmetric distribution of matter with adiabatic flow, and with an EoS in the form of dust cloud, initially at rest to a Schwarzschild black hole. Years later, with the solution of the field equations for a null fluid presented by \cite{Vaidya} and with the junction conditions deduced by \cite{Santos}, it was possible to build more realistic gravitational collapse scenarios for isotropic fluids including dissipative fluxes such as heat flow \citep{OliveiraSantKolass, HerreraDenmatSantos, BonnorOliveiraSant, HerreraPO, Ivanov2012} and shear and bulk viscosities \citep{ChanHerreraSantos}. In addition, models of gravitational collapse for anisotropic fluids with dissipative processes \citep{Chan2001, Govender2019} and even with electric charge \citep{Pinheiro2013, 1Ivanov2019, 2Ivanov2019, Mahomed2019} have been developed in recent years. 

The purpose of this paper is perform a complete analysis on the stability and dynamical collapse of neutron stars with realistic EoSs. In this respect, we first solve the TOV equations that describe the background of each equilibrium configuration. To verify if such configurations are stable or unstable, we proceed to solve the equations that govern the radial oscillations and we calculate the frequencies of the normal modes. Once we are certain that a star is unstable with respect to radial perturbations, we study the temporal evolution of gravitational collapse from the initial static configuration to the formation of an event horizon. During the dissipative gravitational collapse we assume that the unstable star consists of an isotropic fluid with bulk viscosity, a radial heat flow, and an outgoing flux of radiation. Within the context of the classical irreversible thermodynamics we calculate a temperature profile for the case of thermal neutrino transport.

The plan of the present paper is as follows. In Sec. \ref{sec:level2} we present the equations for stellar structure and for radial oscillations. In Sec. \ref{sec:level3} we define the quantities that describe the collapsing fluid, and we derive the junction conditions as well as dynamical equations for unstable compact stars whose final state is the formation of a black hole. Also in Sec. \ref{sec:level3}, we calculate the thermal evolution of the collapsing system and we provide the energy conditions for our model. In Sec. \ref{sec:level4} we present a discussion of the numerical results. Finally, in Sec. \ref{sec:level5} we summarize our conclusions. It is worth mentioning that in all this work we use physical units.


\section{Stellar structure and stability}\label{sec:level2}

In GR the field equations are given by 
\begin{equation}\label{1}
    R_{\mu\nu} - \frac{1}{2}g_{\mu\nu}R = \kappa T_{\mu\nu} ,
\end{equation}
where $\kappa = 8\pi G/c^4$, $R_{\mu\nu}$ is the Ricci tensor, and $R$ denotes the scalar curvature. Here $G$ is the gravitational constant and $c$ is the speed of light in physical units.

In order to study non-rotating stars, we consider a spherically symmetric system whose spacetime is described by the usual line element 
\begin{equation}\label{2}
    ds^2 = g_{\mu\nu}dx^\mu dx^\nu = -e^{2\nu}(dx^0)^2 + e^{2\lambda}dr^2 + r^2d\Omega^2 ,
\end{equation}
where $x^0 = ct$, and $d\Omega^2 = d\theta^2 + \sin^2\theta d\phi^2$ is the line element on the unit 2-sphere. The functions $\nu$ and $\lambda$, in principle, depend on $x^0$ and $r$. 

On the other hand, with respect to the matter-energy distribution, we assume that the system is composed of an isotropic perfect fluid (pf), this is, 
\begin{equation}\label{3}
    T_{\mu\nu}^{\text{pf}} = (\epsilon + p)u_\mu u_\nu + pg_{\mu\nu} ,
\end{equation}
being $u_\mu$ the four-velocity of the fluid, $\epsilon = c^2\rho$ is the energy density (where $\rho$ indicates mass density), and $p$ is the pressure.


\subsection{Background and TOV equations}\label{subsec:level2.1} 

For a star in state of hydrostatic equilibrium none of the quantities depends on the temporal coordinate $x^0$. Consequently, from equations (\ref{1})-(\ref{3}) together with the four-divergence of the energy-momentum tensor (\ref{3}), the stellar structure in hydrostatic equilibrium is governed by TOV equations:
\begin{equation}\label{4}
\dfrac{dm}{dr} = \dfrac{4\pi r^2}{c^2}\epsilon ,
\end{equation}
\vspace*{-0.3cm}
\begin{equation}\label{5}
\dfrac{dp}{dr} = -G \left[ \dfrac{p + \epsilon}{c^2} \right]\left[ \dfrac{m}{r^2} + \dfrac{4\pi rp}{c^2} \right] \left[ 1 - \dfrac{2Gm}{c^2r} \right]^{-1} ,
\end{equation}
\vspace*{-0.3cm}
\begin{equation}\label{6}
\dfrac{d\nu}{dr} = -\dfrac{1}{p + \epsilon}\dfrac{dp}{dr} , 
\end{equation}\\
where $m(r)$ characterizes the mass contained within a sphere of radius $r$. The metric function $\lambda(r)$ is obtained by means of the relation
\begin{equation}\label{7}
    e^{-2\lambda} = 1 - \frac{2Gm}{c^2r} .
\end{equation}
As usual, we define the stellar surface when the matter pressure vanishes, i.e., $p(r = R) = 0$, and the total gravitational mass of the star is given by $M = m(R)$. 

Given an EoS $p = p(\epsilon)$, equations (\ref{4}) and (\ref{5}) can be integrated for a given central density and by guaranteeing regularity in the center of the star, this is,
\begin{align}\label{8}
m(0) &= 0 ,   &    \epsilon(0)   &= \epsilon_c .
\end{align}
In addition, since the equilibrium system is a spherically symmetric star, the exterior spacetime of the star is described by the Schwarzschild metric. Thus, the continuity of the metric on the surface, imposes the following boundary condition for equation (\ref{6})
\begin{equation}\label{9}
    \nu(R) = \frac{1}{2}\ln\left[ 1 - \frac{2GM}{c^2R} \right] .
\end{equation}


\subsection{Equation of state (EoS)}\label{subsec:level2.2} 

In order to study the structure of neutron stars, we need to know its EoS, which is understood as a thermodynamic relation among the energy density, pressure of the fluid, and possibly other local fluid variables such as temperature, baryon density, entropy per baryon, etc. In other words, the EoS of dense matter is a crucial input to close the system of equations (\ref{4})-(\ref{6}). 

Based on theoretical calculations (such as many-body calculations, relativistic mean-field calculations, etc.) the EoSs are usually given in the form of tables. Then, it is common to perform numerical interpolations between the tabulated points. However, since there are different methods and orders of interpolation, this can generate ambiguities in the calculated parameters for neutron stars. Therefore, here we are going to use smooth analytical functions for the EoSs deduced by the authors in \citep{HaenselPotekhin, PotekhinFantinaCPG}, so that numerical errors can be reduced.

\cite{DouchinHaensel} calculated an ``unified EoS'' for neutron stars, describing both the thin crust (composed of electrons and atomic nuclei as well as of free neutrons) and the massive liquid core (which contains electrons (e), neutrons (n), protons (p), muons ($\mu$), and possibly other elementary particles like hyperons, or quark matter). This EoS is based on the effective nuclear interaction SLy (Skyrme Lyon). Years later, \cite{HaenselPotekhin} obtained an analytical representation for the SLy EoS through the following parameterization
\begin{align}\label{10}
\mathcal{P}(\xi) =& \dfrac{a_1 + a_2\xi + a_3\xi^3}{1 + a_4\xi}K_0(a_5(\xi - a_6))  \nonumber  \\
&+ (a_7 + a_8\xi)K_0(a_9(a_{10} - \xi))   \nonumber  \\
&+ (a_{11} + a_{12}\xi)K_0(a_{13}(a_{14} - \xi))   \nonumber  \\
&+ (a_{15} + a_{16}\xi)K_0(a_{17}(a_{18} - \xi)) ,
\end{align}
where $\mathcal{P} = \log(p/ \text{dyn}\ \text{cm}^{-2})$, $\xi = \log(\rho/ \text{g}\ \text{cm}^{-3})$, and \ $K_0(x) = 1/(e^x +1)$. The coefficients of this fit $a_i$ can be found in \citep{HaenselPotekhin}. Note that we are using ``$\log$'' to denote ``$\log_{10}$'', while the natural logarithm will be  denoted by ``$\ln$''. Such an EoS has an important feature; it is compatible with the recent LIGO-Virgo constraints \citep{Abbott}.

\begin{figure}
 \includegraphics[width=\columnwidth]{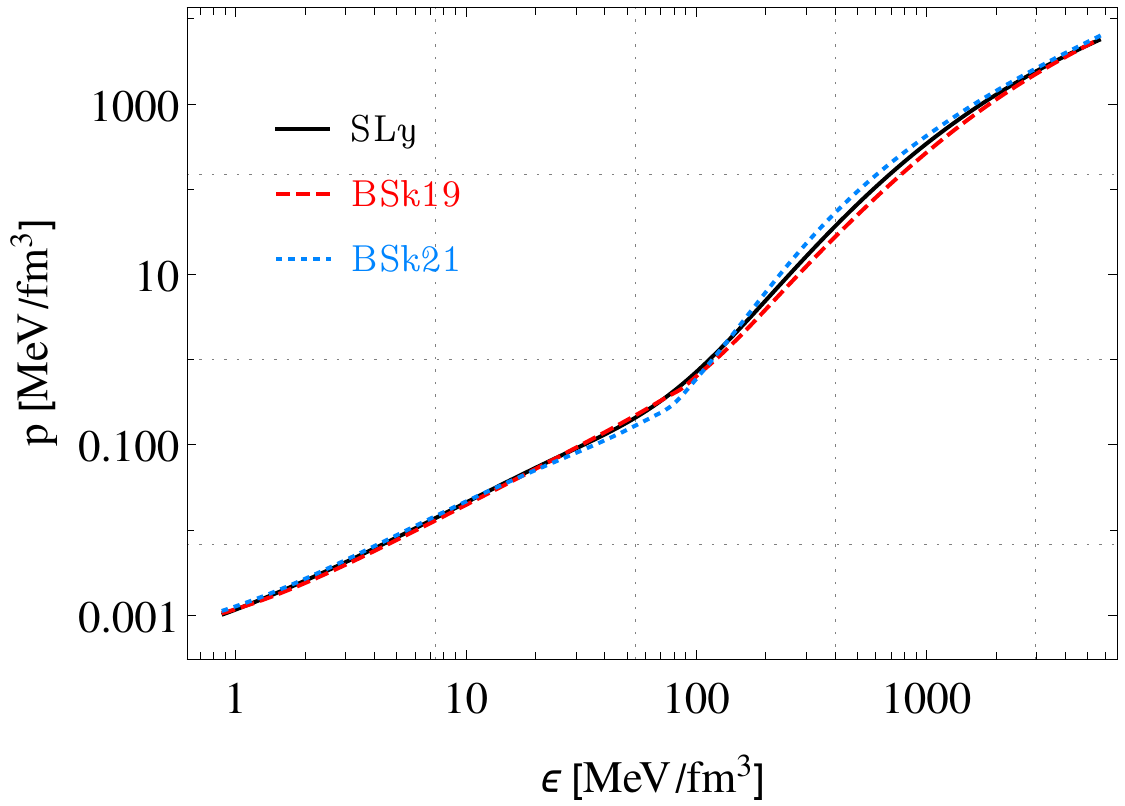}
 \caption{SLy, BSk19, and BSk21 equations of state, given by (\ref{10}) and (\ref{11}).}  
 \label{figure1} 
\end{figure}

By completeness, we will also use two of the unified Brussels-Montreal EoSs which are based on the nuclear energy-density functionals with generalized Skyrme effective forces, known as BSk19 and BSk21 \citep{PotekhinFantinaCPG}, and which significantly differ by stiffness. In these cases, the analytical parametrization is given by 
\begin{align}\label{11}
\mathcal{P}(\xi) =& \dfrac{a_1 + a_2\xi + a_3\xi^3}{1 + a_4\xi} \lbrace\exp[a_5(\xi-a_6)]+ 1\rbrace^{-1}  \nonumber  \\
&+ (a_7 + a_8\xi)\lbrace \exp[a_9(a_6 - \xi)]+ 1 \rbrace^{-1}  \nonumber  \\
&+ (a_{10} + a_{11}\xi)\lbrace \exp[a_{12}(a_{13} - \xi)]+ 1 \rbrace^{-1}  \nonumber  \\
&+ (a_{14} + a_{15}\xi)\lbrace \exp[a_{16}(a_{17} - \xi)]+ 1 \rbrace^{-1}  \nonumber  \\
&+ \frac{a_{18}}{1 + [a_{19}(\xi - a_{20})]^2} + \frac{a_{21}}{1 + [a_{22}(\xi - a_{23})]^2} , 
\end{align}
whose coefficients $a_i$ are given in \citep{PotekhinFantinaCPG}. The EoSs (\ref{10}) and (\ref{11}) are showed in figure \ref{figure1}, and will be used in this paper in order to solve  numerically the stellar structure equations.

\subsection{Stability criteria}\label{subsec:level2.3} 

Along the sequence of equilibrium configurations on the mass-central density curve, a necessary but not sufficient condition for stellar stability is \citep{Glendenning}: 
\begin{equation}\label{12}
    \frac{\partial M(\rho_c)}{\partial\rho_c} = 0 ,
\end{equation}
this is, stars consisting of a perfect fluid (\ref{3}) can pass from stability to instability with respect to some radial normal mode only at a value of central density for which the equilibrium mass is stationary.  

Nevertheless, a sufficient condition for the analysis of stability of relativistic stars with respect to radial perturbations is determine the frequencies of normal modes of vibration. The equations that govern the radial adiabatic oscillations for a spherical star with isotropic pressures  in GR were first derived by  \cite{Chandrasekhar, Chandrasekhar1}, where Einstein equations are linearized around the equilibrium configuration to generate a Sturm-Liouville problem. Thenceforth, equations that govern such pulsations have been rewritten in several forms in the literature \citep{Chanmugam, VathChanmugam, Gondek, KokkotasRuoff, PanotopoulosLopes2017, Sagun2020}, some of them being suitable for numerical computations. Here we will use the Gondek's form \citep{Gondek}. If it is assumed that the radial displacement function (as well as the metric functions and thermodynamics quantities) has a harmonic time dependence like $\delta r(t, r) = \varsigma(r) e^{i\omega t}$, and by defining a new variable $\zeta = \varsigma/r$, we have two first-order time-independent equations
\begin{equation}\label{13}
    \dfrac{d\zeta}{dr} = -\dfrac{1}{r}\left( 3\zeta + \dfrac{\Delta p}{\gamma p} \right) - \dfrac{dp}{dr}\dfrac{\zeta}{p+ \epsilon} ,
\end{equation}
\begin{align}\label{14}
    \dfrac{d(\Delta p)}{dr} = & \ \zeta\left[ \dfrac{\omega^2}{c^2}e^{2(\lambda - \nu)}(p + \epsilon)r - 4\dfrac{dp}{dr} \right.  \nonumber   \\
    & \left.- \kappa e^{2\lambda}(p + \epsilon)rp + \dfrac{r}{p + \epsilon}\left( \dfrac{dp}{dr} \right)^2  \right]   \nonumber  \\
    & + \Delta p\left[ \dfrac{1}{p+ \epsilon}\dfrac{dp}{dr} - \frac{\kappa}{2} (p + \epsilon)re^{2\lambda} \right] ,
\end{align}
where $\gamma = (1 + \epsilon/p)dp/d\epsilon$ denotes the \textit{adiabatic index} at constant specific entropy,  and $\Delta p = \delta p + \varsigma dp/dr$ is the Lagrangian perturbation of the pressure, with $\delta p$ being the Eulerian perturbation \citep{ShapiroT}.

Equation (\ref{13}) has a singularity at the origin ($r = 0$). In order that $d\zeta/dr$ to be regular everywhere, it is required that as $r \rightarrow 0$ the coefficient of $1/r$ term must vanish, so that
\begin{equation}\label{15}
    \Delta p = -3\gamma\zeta p  \qquad \ \  \text{as}  \qquad  \ \  r\rightarrow 0 .
\end{equation}
The surface of the star is determined by the condition that for $r\rightarrow R$, we must have $p \rightarrow 0$. This implies a condition on the Lagrangian perturbation of the pressure on the stellar surface, 
\begin{equation}\label{16}
    \Delta p = 0  \qquad \ \  \text{as}  \qquad  \ \  r\rightarrow R .
\end{equation}


\section{Gravitational collapse}\label{sec:level3}

\subsection{Interior spacetime and transport equations}\label{subsec:level3.1} 

It is well known that the formation of neutron stars and black holes is the result of a period of radiating gravitational collapse in which massless particles (photons and neutrinos) carry thermal energy for exterior spacetime \citep{HerreraSantos, Mitra}. In fact, the gravitational collapse is a highly dissipative process, so it becomes mandatory to take into account terms that describe departure from the equilibrium state (characterised by the absence of transport phenomena) \citep{HerreraPO}. Therefore, it is relevant to invoke the relativistic hydrodynamic equations of non-perfect fluids (npf). The dissipative contribution to the energy-momentum tensor can be written into two parts as
\begin{equation}\label{GC1}
    T_{\mu\nu}^{\text{diss}} = \mathcal{S}_{\mu\nu} + T_{\mu\nu}^{\text{flux}} ,
\end{equation}
where $\mathcal{S}_{\mu\nu}$ represents the viscous contributions and is usually known like \textit{viscous stress tensor}, given by
\begin{equation}\label{GC2}
    \mathcal{S}_{\mu\nu} = \pi_{\mu\nu} + \Pi h_{\mu\nu} ,
\end{equation}
being $\pi_{\mu\nu}$ the shear stress tensor, $\Pi$ the bulk viscous pressure, and $h_{\mu\nu} = g_{\mu\nu} + u_\mu u_\nu$ is the standard projection tensor orthogonal to the four-velocity.

On the other hand, by recalling that in the Newtonian formalism the heat flow is described by a three-vector $\vec{q} = -\bar{\kappa}\nabla T$, the relativistic generation of energy fluxes is given by $T_{\mu\nu}^{\text{flux}} = 2q_{(\mu} u_{\nu)}$. Parentheses around the set of indices denote symmetrization and $q_\mu$ is the heat flux four-vector. Ultimately, by collecting all these contributions, the energy-momentum tensor of a non-perfect fluid $T_{\mu\nu}^{\text{npf}} = T_{\mu\nu}^{\text{pf}} + T_{\mu\nu}^{\text{diss}}$ can be written explicitly as
\begin{equation}\label{GC3}
    T_{\mu\nu}^{\text{npf}} = \epsilon u_\mu u_\nu + (p + \Pi)h_{\mu\nu} + q_\mu u_\nu + q_\nu u_\mu + \pi_{\mu\nu} ,
\end{equation}
where $\Pi$, $q_\mu$ and $\pi_{\mu\nu}$ are also known as thermodynamic fluxes and they account for the deviations of the dissipative fluid from a perfect fluid. They satisfy the following properties
\begin{align}\label{GC4}
    u_\mu q^\mu &= 0 ,   &    u_\mu \pi^{\mu\nu} &= 0 ,   &     \pi_\mu^{\ \mu} &= 0 .
\end{align}

Within the formalism of Classical Irreversible Thermodynamics (which in the last decades has been notoriously criticized by non-causality and instability but is still widely used as an approximation), from relativistic-hydrodynamic equations and laws of thermodynamics, the transport equations for the propagation of dissipative fluxes are given by \citep{Rezzolla}
\begin{subequations}
 \begin{align}
    \Pi &= -\bar{\zeta} \Theta ,  \label{GC5a}  \\
    q_\mu &= -\frac{\bar{\kappa}}{c} \left(h_\mu^{\ \nu}\nabla_\nu T + Tu^\nu\nabla_\nu u_\mu \right) ,   \label{GC5b}  \\
    \pi_{\mu\nu} &= -2\eta \sigma_{\mu\nu} ,   \label{GC5c}
 \end{align}
\end{subequations}
where $\bar{\zeta}$, $\bar{\kappa}$ and $\eta$ are the thermodynamic coefficients of bulk viscosity,  thermal conductivity and shear viscosity, respectively. The heat flux and the temperature profile ($T$) inside the star are related through the heat conduction equation (\ref{GC5b}), this is, the relativistic Fourier law \citep{Eckart}.

Our proposal here is to introduce a temporal dependency on the metric components in such a way that under a certain limit we can fall into the static case. Thus, in order to perform a realistic time description about the dynamic collapse of neutron stars, we assume that interior spacetime of the collapsing star is described by the line element 
\begin{equation}\label{GC6}
    ds^2 = -e^{2\nu(r)}(dx^0)^2 + e^{2\lambda(r)}f(x^0)dr^2 + r^2f(x^0)d\Omega^2 ,
\end{equation}
in comoving coordinates. In the static limit $f(x^0) \rightarrow 1$ we recover equation (\ref{2}) which describes the initial static star. Once we know that a certain configuration in hydrostatic equilibrium is unstable (with its ultimate fate being the gravitational collapse), the dynamical instability is governed by the metric (\ref{GC6}). Taking into account that the dimensionless four-velocity satisfies the normalisation condition (i.e., $u_\mu u^\mu =-1$), and the first relation in (\ref{GC4}), we have 
\begin{align}\label{GC7}
    u^\mu &= e^{-\nu}\delta_0^\mu ,   &    q^\mu &= \frac{q}{c}\delta_r^\mu ,
\end{align}
where $q = q(x^0, r)$ is the rate of energy flow per unit area along the radial coordinate. 

In addition, it is useful to introduce some quantities that describe the kinematic properties of the collapsing fluid: the four-acceleration, the expansion scalar, and the shear tensor, namely, 
\begin{subequations}
 \begin{align}
    a^\mu &= c^2u^\alpha\nabla_\alpha u^\mu = \frac{c^2\nu'}{e^{2\lambda}f}\delta_r^\mu ,  \label{GC8a}  \\
    \Theta &= c\nabla_\mu u^\mu = \frac{3c}{2e^\nu}\frac{\dot{f}}{f} ,  \label{GC8b}  \\
    \sigma_{\mu\nu} &= c\nabla_{(\mu}u_{\nu)} + \frac{1}{c}a_{(\mu}u_{\nu)} - \frac{1}{3}\Theta h_{\mu\nu} = 0 ,  \label{GC8c}
 \end{align}
\end{subequations}
where the dot and the prime stand for partial derivative with respect to $x^0$ and $r$, respectively. It is evident that for this model we have  $\pi_{\mu\nu} =0$.


\subsection{Junction conditions and dynamical equations on the stellar surface}\label{subsec:level3.2} 

As the non-perfect fluid collapses, it emits radiation in the form of a null fluid, hence the exterior spacetime is described by the Vaidya metric \citep{Vaidya, Mkenyeleye}
\begin{equation}\label{GC9}
    ds^2 = -\left[ 1 - \frac{2Gm(v)}{c^2\chi} \right]c^2dv^2 + 2\varepsilon c dvd\chi + \chi^2d\Omega^2 ,
\end{equation}
where $m(v)$ represents the mass function that depends on the retarded time $v$, and $\varepsilon = \pm 1$ describes the incoming (outgoing) flux of radiation, respectively. In our case the radiation is expelled into outer region so that $dm/dv \leq 0$. In other words, the collapsing star is losing mass due to the emitted radiation.

During the gravitational collapse the interior spacetime is described by (\ref{GC6}), while the outer region is described by the line element (\ref{GC9}). Therefore, in order to guarantee continuity and smoothness between these geometries, it is necessary to invoke the junction conditions on the boundary surface connecting the two regions (a spherical hyper-surface usually denoted by $\Sigma$) established by \cite{Israel1, Israel2}. Such conditions lead to the following relations
\begin{subequations}
 \begin{align}
    \chi_\Sigma &= \left[r\sqrt{f}\right]_\Sigma ,  \label{GC10a}  \\
    m_\Sigma &= \frac{c^2R\sqrt{f}}{2G}\left[ 1 + \frac{r^2}{4e^{2\nu}}\frac{\dot{f}^2}{f} - \frac{1}{e^{2\lambda}} \right]_\Sigma ,   \label{GC10b}  \\
    p_\Sigma &= \left[ \frac{q}{c}e^{\lambda}\sqrt{f} + \bar{\zeta}\Theta \right]_\Sigma ,   \label{GC10c}   \\
    z_\Sigma &= \left[ \frac{dv}{d\tau} \right]_\Sigma - 1 = \left[ \frac{1}{e^\lambda} + \frac{r}{2e^\nu}\frac{\dot{f}}{\sqrt{f}} \right]_\Sigma^{-1} - 1 ,   \label{GC10d}
\end{align}
\end{subequations}
being $z_\Sigma$ the boundary redshift of the radial radiation emitted by the collapsing star, $\tau$ is the proper time defined on $\Sigma$ and $d\tau = e^{\nu(R)}dt$. It is clearly seen that in the static limit $m_\Sigma$ corresponds to the total mass of the initial Schwarzschild configuration $M$, and the redshift is reduced to $z_\Sigma = e^{\lambda(R)} - 1$, this is, the gravitational redshift of light emitted at the surface of the neutron star initially in hydrostatic equilibrium.

Now we can obtain the total luminosity perceived by an observer at rest at infinity, namely,
\begin{equation}\label{GC11}
    L_\infty = -c^2\left[ \frac{dm}{dv} \right]_\Sigma = 4\pi f^{3/2}\left[ \frac{q r^2}{e^\lambda}\left( \frac{re^\lambda}{2e^\nu}\frac{\dot{f}}{\sqrt{f}} + 1 \right)^2 \right]_\Sigma ,
\end{equation}
so that the luminosity of the collapsing sphere as measured on its surface is related with $L_\infty$ through equation
\begin{equation}\label{GC12}
   L_\Sigma = L_\infty(1 + z_\Sigma)^2 = 4\pi f^{3/2}\left[ qr^2e^{\lambda} \right]_\Sigma .
\end{equation}


\subsection{Field equations}\label{subsec:level3.3} 

For the energy-momentum tensor (\ref{GC3}) and line element (\ref{GC6}), the field equations (\ref{1}) assume the explicit form
\begin{align}
    \epsilon(x^0, r) =& \frac{1}{\kappa}\left[ -\frac{1 - e^{2\lambda} - 2r\lambda'}{r^2e^{2\lambda}f} + \frac{3}{4e^{2\nu}}\frac{\dot{f}^2}{f^2} \right] ,    \label{GC13}   \\
    p(x^0, r) =& \frac{1}{\kappa}\left[ \frac{1 - e^{2\lambda} + 2r\nu'}{r^2e^{2\lambda}f} - \frac{1}{e^{2\nu}}\left( \frac{\ddot{f}}{f} - \frac{\dot{f}^2}{4f^2} \right) \right] + \frac{3c\bar{\zeta}}{2e^\nu}\frac{\dot{f}}{f} ,    \label{GC14}   \\
    q(x^0, r) =& -\frac{c \nu'}{\kappa e^{\nu + 2\lambda}}\frac{\dot{f}}{f^2} ,    \label{GC15}
\end{align}
that is, equations (\ref{GC13}), (\ref{GC14}) and (\ref{GC15}) describe the temporal and radial behavior of the energy density, pressure and heat flow during gravitational collapse for each unstable stellar configuration.

In view of this, by considering equations (\ref{GC14}) and (\ref{GC15}) into (\ref{GC10c}) on the stellar surface, the temporal evolution of our model is fixed by
\begin{equation}\label{GC16}
    \frac{d^2f}{dt^2} - \frac{1}{4f}\left( \frac{df}{dt} \right)^2 - \frac{GM}{cR^2}\frac{1}{\sqrt{f}}\frac{df}{dt} = 0 ,
\end{equation}
so the first integration of (\ref{GC16}) yields
\begin{equation}\label{GC17}
    \frac{df}{dt} = \frac{4GM}{cR^2}\left[ \sqrt{f} - f^{1/4} \right] ,  
\end{equation}
whose solution is given by
\begin{equation}\label{GC18}
    t = \frac{cR^2}{2GM}\left[ \sqrt{f} + 2f^{1/4} + 2\ln\left(1 - f^{1/4}\right) \right] .
\end{equation}

The divergence of the expression (\ref{GC10d}) indicates the formation of an event horizon (i.e., the surface that encloses the interior spacetime of a black hole and from which no information can escape) resulting from gravitational collapse of an unstable star,
\begin{equation}\label{GC19}
    f_{bh} = 16\left[ \frac{GM}{c^2R} \right]^4 .
\end{equation}
For configurations in gravitational collapse we must have $0 < f \leq 1$, this is, the function $f$ decrease monotonically from the value $f = 1$ (when the model is static) to $f = f_{bh}$ (when the star becomes a black hole). The time goes from $t = -\infty$ to $t = t_{bh}$, but a time displacement can be done without loss of generality. Equation (\ref{GC18}) provides $t$ as a function of $f$, however, it is more useful to obtain $f(t)$ in order to analyze the physical quantities as a function of time as a star collapses. Taking this into account, it is convenient to numerically solve equation (\ref{GC16}) as if it were an final value problem by specifying a value of $f(t)$ and $df(t)/dt$ at time $t = t_{bh}$. These two final conditions are established through equations (\ref{GC17})-(\ref{GC19}).

It is worth emphasizing that when $f \rightarrow 1$ the external solution is Schwarzschild-type, the hydrostatic equilibrium is governed by the TOV equations, and all dissipative flows are null. However, when the star gradually begins to collapse the thermodynamic quantities are described by the metric functions in (\ref{GC6}), and the exterior Vaidya spacetime is completely determined by $m(v)$ which, in turn, is determined from the interior metric through the solution of equation (\ref{GC16}).

The mass corresponding to the black hole formed is obtained from equations (\ref{GC10b}), (\ref{GC17}) and (\ref{GC19}),
\begin{equation}\label{GC20}
    m_{bh} = 2\frac{GM^2}{c^2R} ,
\end{equation}
which depends only on the initial conditions of the unstable neutron star. In fact, such initial data depends mainly on the central density and the EoS adopted to describe the unstable equilibrium. In our analysis, what defines dynamical instability with respect to radial perturbations is the imaginary frequency of the lowest mode of oscillation. As pointed out by \cite{JoshiGoswami}, the initial data play an important role in causing a black hole (where the spacetime singularity is necessarily hidden behind the event horizon according to the cosmic censorship conjecture) or a naked singularity (when the singularity is not covered by an event horizon) in the final state of a gravitational collapse.


\subsection{Thermal evolution}\label{subsec:level3.4} 

In astronomy and astrophysics, the stars are often modelled like blackbodies in order to determine its surface temperature \citep{Vitense}. Under this approximation, if we assume that the collapsing star radiates as a blackbody, we can calculate an effective surface temperature measured by an observer at rest at infinity by means of the Stefan-Boltzmann Law, namely 
\begin{equation}\label{GC21}
    T_{\text{eff}, \Sigma}^4 = \frac{L_\infty}{4\pi \sigma fR^2} ,
\end{equation}
where $\sigma = 5.6704 \times 10^{-8}\ \rm W \cdot m^{-2} \cdot K^{-4}$ is the  Stefan's constant. As a result, according to equations (\ref{GC11}), (\ref{GC15}), and (\ref{GC17}), we get
\begin{equation}\label{GC22}
    T_{\text{eff}, \Sigma}^4 = \frac{cGM^2}{2\pi\sigma R^4} \left[ \frac{f^{-1/4} -1}{f} \right] \left[ 1 - \frac{2GM(f^{-1/4}-1)}{c^2R - 2GM} \right]^2 .
\end{equation}

Let us proceed to examine the thermodynamical behavior of the collapsing dissipative fluid and their implications. Field equations do not give us information on how the temperature of a star evolves as it collapses. Accordingly, we have to resort to relativistic transport equations. In particular, the Fourier-Eckart law (\ref{GC5b}) for heat transport takes the form 
\begin{equation}\label{GC23}
    \bar{\kappa}\frac{\partial}{\partial r}( e^\nu T ) = -e^{\nu + 2\lambda}fq ,
\end{equation}
and for physically reasonable models, it is often assumed that heat is carried to exterior spacetime through thermally generated neutrinos \citep{Martinez, GovenderMM1998, GovenderMM1999}, so that the thermal conductivity can be written as  $\bar{\kappa} = \alpha T^{3 - \beta}$. Thus, from the expression (\ref{GC23}) we obtain the following differential equation for temperature
\begin{equation}\label{GC24}
    \frac{\partial}{\partial r}\left[ (e^\nu T)^4 \right] + \frac{4}{\alpha}fq e^{(4-\beta)\nu + 2\lambda}\left[ (e^\nu T)^4 \right]^{\beta/4} = 0 ,
\end{equation}
which entails that 
\begin{equation}\label{GC25}
    (e^\nu T)^{4-\beta} = \frac{\beta - 4}{\alpha}\int fqe^{(4-\beta)\nu + 2\lambda}dr + \mathcal{C}(t) ,  \quad  \beta \neq 4 .
\end{equation}

The integration function $\mathcal{C}(t)$ can be determined from the effective surface temperature (\ref{GC22}). In other words, we use a boundary condition for temperature profile, given by $T_\Sigma = T_{\text{eff}, \Sigma}$. Therefore, taking into account equations (\ref{GC15}), (\ref{GC17}) and (\ref{GC22}), the integral in (\ref{GC25}) can be explicitly calculated to yield
\begin{align}\label{GC26}
    T^{4 - \beta}(x^0, r) =& \ \frac{c^3M}{2\pi\alpha R^2}\left( \frac{\beta-4}{\beta-3} \right) \left[ \frac{f^{1/2} - f^{1/4}}{f} \right]  \nonumber  \\  
    &\times \left[ \frac{1}{e^\nu} - \frac{1}{a}\left( \frac{a}{e^\nu} \right)^{4-\beta} \right] + \left[\frac{cGM^2( f^{-1/4}-1 )}{2\pi\sigma R^4f} \right]^{\frac{4 - \beta}{4}}  \nonumber  \\
    &\times \left[ \frac{a^2}{e^{2\nu}} - \frac{2GM(f^{-1/4} - 1)}{c^2Re^{2\nu}} \right]^{\frac{4-\beta}{2}} ,
\end{align}
for $\beta \neq 3,4$, and where $a = e^{\nu(R)}$, i.e. a constant defined on the surface of the star.

The surface temperature given by (\ref{GC22}) is of the order of $10^{12}\ \rm K$ for unstable neutron stars and we expect it to rise as we approach the center. In addition, it is well known \citep{ShapiroT} that at very high temperatures ($T\gtrsim 10^9\ \rm K$) the dominant mode of energy loss via neutrinos in collapsing stars is from the so-called Urca process. The simplest Urca process involves neutrons, protons and electrons:
\begin{equation}\label{GC27}
   \rm n \rightarrow p + e + \bar{\nu}_e ,   \qquad    e + p \rightarrow n + \nu_e ,
\end{equation}
this is, a chain of two (direct and inverse) reactions, where $\nu_e$ and $\bar{\nu}_e$ are the electron neutrino and antineutrino, respectively. Thus, since the thermal neutrino processes are relevant during the gravitational collapse, \cite{Martinez} assumed that neutrinos are thermally generated with energies close to $k_BT$ in order to bring about the following effective mean free path of radiation
\begin{equation}\label{GC28}
    \lambda_{\rm eff} \sim \sqrt{\lambda_n \lambda_e} \approx \frac{ 80.9\ \rm kg \cdot J^{3/2} \cdot m^{-2} }{ \rho \sqrt{\mathcal{Y}_e}(k_BT)^{3/2} } ,
\end{equation}
being $\rho$ the mass density in $\rm kg/m^3$, $\mathcal{Y}_e$ the electron fraction, and $k_B = 1.38 \times 10^{-23}\ \rm J \cdot K^{-1}$ is the Boltzmann's constant.

If the collapsing system is a mixture of matter and radiation, the thermal conductivity coefficient and the bulk viscous coefficient are given by \citep{Weinberg}
\begin{align}
    \bar{\kappa} &= \frac{4b}{3} T^3\lambda_{\rm eff} ,   \label{GC29}   \\
    \bar{\zeta} &= \frac{4b}{c^2}T^4\lambda_{\rm eff}\left[ \frac{1}{3} - \left( \frac{\partial p}{\partial \epsilon} \right)_n \right]^2 ,  \label{GC30}
\end{align}
where $b= 7\sigma/8$ for neutrinos. Then we can identify $\beta = 3/2$, and for $\mathcal{Y}_e =0.2$ the proportionality factor $\alpha$ takes the form
\begin{equation}\label{GC31}
    \alpha = \frac{ 2.3\times 10^{29}}{\rho} \frac{\rm kg \cdot W}{\rm m^4 \cdot K^{5/2}}\ .
\end{equation}

Nonetheless, for our model the approximate expression given in (\ref{GC28}) leads to $\Pi \sim 10^{37}\ \rm Pa$ and, therefore, a negative pressure according to equation (\ref{GC14}). For this reason, we decide to use the effective mean free path obtained by \cite{ShapiroT}, namely
\begin{equation}\label{GC32}
    \lambda_{\rm eff} \sim  \frac{ 1.3\times 10^{-6}\ \rm m \cdot (kg/m^3)^{7/6} \cdot J^{5/2} }{ \rho^{7/6} (k_BT)^{5/2} } .
\end{equation}
We would like to have a more exact expression for $\lambda_{\rm eff}$, but for our purposes it is enough to adopt the form (\ref{GC32}). The choice of this mean free path, which admittedly provides only an approximated description, is justified because the main aim here is to test the validity of our model. Under such an approximation, we have $\beta = 5/2$, and 
\begin{equation}\label{GC33}
    \alpha = \frac{ 1.2\times 10^{44}}{\rho^{7/6}} \frac{\rm W \cdot (kg/m^3)^{7/6}}{\rm m \cdot K^{3/2}} .
\end{equation}


\subsection{Energy conditions for the collapsing fluid}\label{subsec:level3.5} 

In order for our stellar collapse model to be physically reasonable, it is necessary that it obey the weak, dominant and strong energy conditions. To verify if such conditions are satisfied, we follow the same procedure used by the authors in \citep{Kolassis_1988}. As the energy-momentum tensor corresponding to a non-perfect fluid has components outside the diagonal, it is required its diagonalization. Indeed, the eigenvalues $\Upsilon$ of the energy-momentum tensor diagonalized are the roots of the equation $\vert T_{\mu\nu}^{\rm npf} - \Upsilon g_{\mu\nu} \vert = 0$, which can be explicitly written as
\begin{equation}\label{GC34}
    \begin{vmatrix}
     (\epsilon + \Upsilon)e^{2\nu}  &  -\tilde{q}\sqrt{f}e^{\nu+\lambda}  &  0  &  0  \\
     -\tilde{q}\sqrt{f}e^{\nu+\lambda}  &  (\mathcal{Q} - \Upsilon)fe^{2\lambda}  &  0  &  0  \\
     0  &  0  &  (\mathcal{Q} - \Upsilon)fr^2  &  0  \\
     0  &  0  &  0  &  (\mathcal{Q} - \Upsilon)fr^2\sin^2\theta 
\end{vmatrix}  = 0 ,
\end{equation}
being $\tilde{q} = \frac{q}{c}\sqrt{f}e^\lambda$ and $\mathcal{Q} = p+\Pi$. Then, equation (\ref{GC34}) has the roots
\begin{subequations}
\begin{align}
    \Upsilon_0 &= -\frac{1}{2}(\epsilon - \mathcal{Q} + \Delta) ,   \\
    \Upsilon_1 &= -\frac{1}{2}(\epsilon - \mathcal{Q} - \Delta) ,   \\
    \Upsilon_2 &= \Upsilon_3 = \mathcal{Q} ,
\end{align}
\end{subequations}
where we have defined $\Delta = \sqrt{ (\epsilon + \mathcal{Q})^2 - 4\tilde{q}^2 }$.

According to Ref. \citep{Kolassis_1988}, the energy conditions in terms of eigenvalues are given by:
\begin{itemize}
    \item[$\star$] \textbf{Weak energy conditions (WEC)} 
    \[
    \begin{cases}
        -\Upsilon_0 \geq 0 ,  \\
        -\Upsilon_0 + \Upsilon_i \geq 0, \qquad   \text{for} \ i = 1, 2, 3.
    \end{cases}
    \]
    This implies the following inequalities
    \begin{subequations}
    \begin{align} 
        &\Delta \geq 0,  \label{GC36a}  \\  
        &\epsilon + \Delta - p - \Pi \geq 0 ,  \label{GC36b}  \\  
        &\epsilon + \Delta + p + \Pi \geq 0 .  \label{GC36c}
    \end{align}
    \end{subequations}

    \item[$\star$] \textbf{Dominant energy conditions (DEC)} 
    \[
    \begin{cases}
        -\Upsilon_0 \geq 0 ,  \\
        -\Upsilon_0 + \Upsilon_i \geq 0, \qquad   \text{for} \ i = 1, 2, 3. \\
        \Upsilon_0 + \Upsilon_i \leq 0, \qquad   \text{for} \ i = 1, 2, 3.
    \end{cases}
    \]
    The first two inequalities have already been considered in weak energy conditions. With respect to the third inequality, we have
    \begin{subequations}
    \begin{align}
        &\epsilon - p - \Pi \geq 0 ,  \label{GC37a}  \\  
        &\epsilon - 3(p + \Pi) + \Delta \geq 0 .   \label{GC37b} 
    \end{align}
    \end{subequations}
    
    \item[$\star$] \textbf{Strong energy conditions (SEC)} 
    \[
    \begin{cases}
        -\Upsilon_0 + \sum_{i=1}^3 \Upsilon_i \geq 0 ,   \\
        -\Upsilon_0 + \Upsilon_i \geq 0, \qquad   \text{for} \ i = 1, 2, 3.
    \end{cases}
    \]
    From the first inequality, we get
    \begin{equation}\label{GC38}
        \Delta + 2(p + \Pi) \geq 0 ,
    \end{equation}
    while the second inequality has been included in the other conditions.
\end{itemize}

If the energy conditions (\ref{GC36a}) and (\ref{GC37a}) are respected, it is clear that inequality (\ref{GC36b}) will be satisfied. Similarly, the satisfaction of the conditions (\ref{GC37a}) and (\ref{GC38}) guarantees the fulfillment of (\ref{GC36c}). Therefore, it is only necessary to verify the energy conditions (\ref{GC36a}), (\ref{GC37a}), (\ref{GC37b}) and (\ref{GC38}) to know if our model is physically acceptable. A graphical analysis thereof will be shown in the next section for an unstable neutron star during its collapse process.


\section{Numerical results and discussion}\label{sec:level4}

\begin{figure*}
 \includegraphics[width=8.6cm]{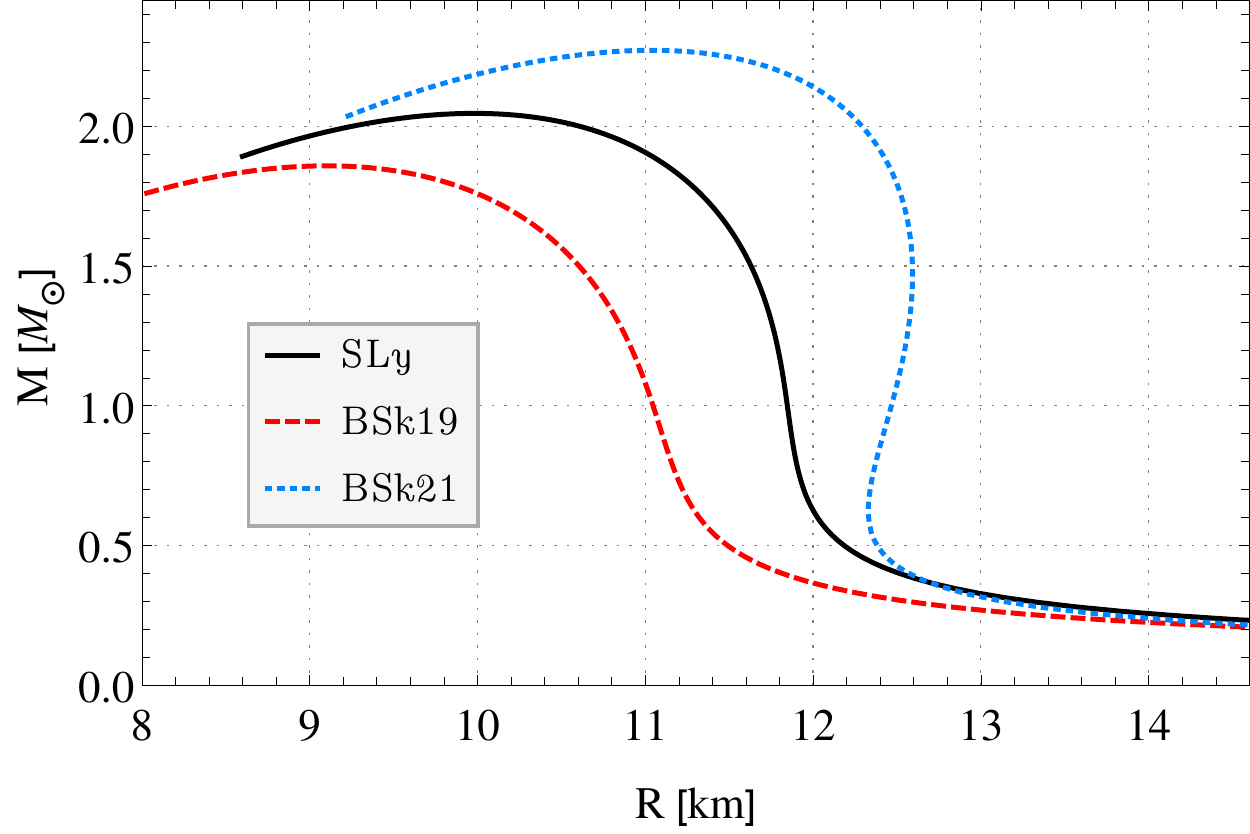} \
 \includegraphics[width=8.6cm]{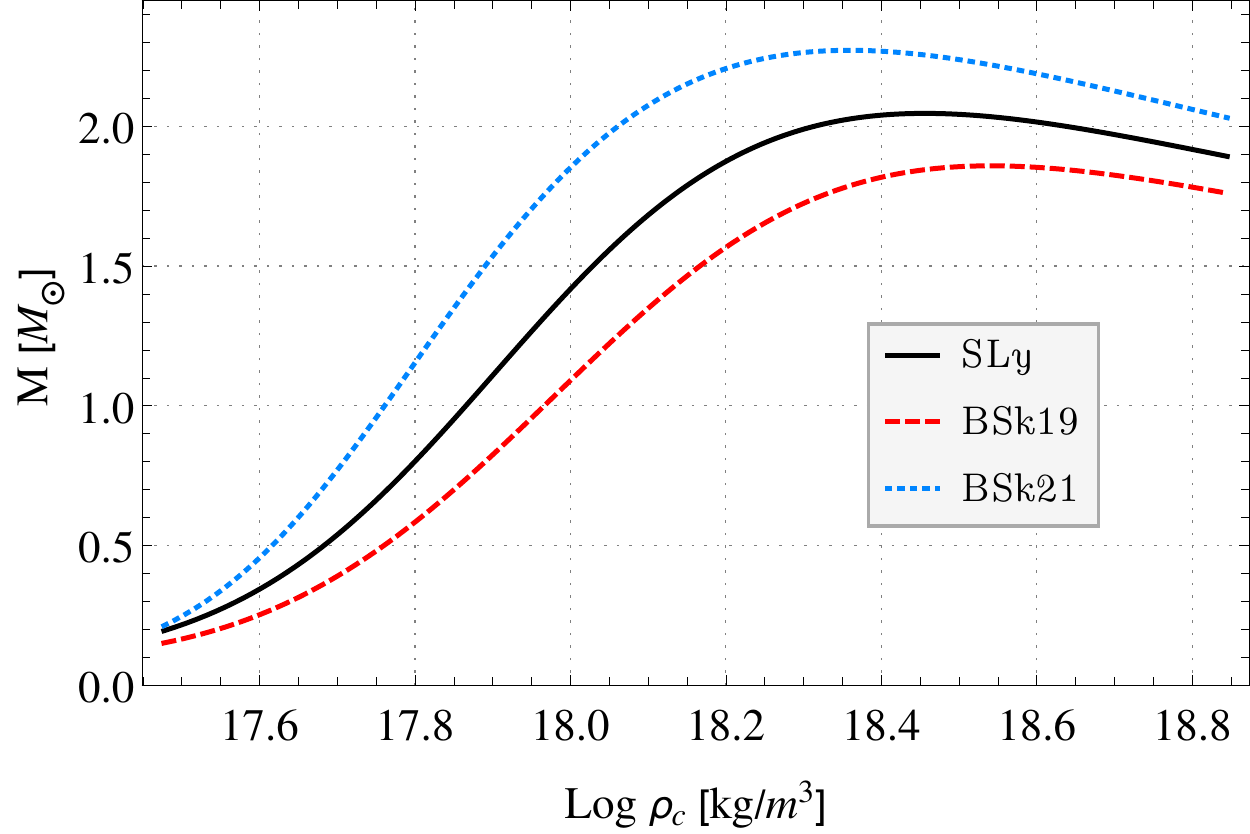}
 \caption{ Mass-radius diagram (left panel) and mass-central density relation (right panel) for neutron stars in GR with SLy, BSk19, and BSk21 equations of state. For each EoS the stable stars are on the region with positive slope on the $M(\rho_c)$ curve.} 
 \label{figure2}
\end{figure*}

Given a value of central mass density $\rho_c$, the background equations (\ref{4})-(\ref{6}) can be numerically solved subject to conditions (\ref{8}) and (\ref{9}). Thus, for each equilibrium configuration with radius $R$ and total gravitational mass $M = m(R)$, we obtain the metric functions $\nu$ and $\lambda$ as well as the thermodynamic quantities as functions of the radial coordinate. The mass-radius diagram of sequences of static stellar configurations with EoSs (\ref{10}) and (\ref{11}) is presented in the left panel of Fig. \ref{figure2}. Our second step is to find out whether these equilibrium configurations are stable or unstable with respect to a radial perturbation. There is a critical central density $\rho_{crit}$ for which the gravitational mass as a function of $\rho_c$ is maximum, so that when $\rho_c = \rho_{crit}$ the frequency of the fundamental mode must vanish. In table \ref{table1} we present these critical densities for each EoS. In the right panel of Fig. \ref{figure2} the critical densities correspond to the maximum of each curve in the mass-central density diagram. Thus, for central densities lower than the critical density (i.e., when $dM/d\rho_c > 0$) all neutron stars belonging to this region are stable. Stellar configurations that do not satisfy this necessary condition for stability are unstable. The first observation of gravitational waves from a binary neutron stars system, GW170817, detected by the Advanced LIGO and Advanced Virgo gravitational-wave detectors \citep{Abbott}, suggests that the mass  for each component is in the range $1.17 < M < 1.6$, in solar mass units. In tables \ref{table2}, \ref{table3}, and \ref{table4} we can see that they are in the stable region of our model, to all hypotheses of equation of state considered. Therefore, our model is compatible with the existence of stable stars in this mass range.

\begin{table}
\centering
\caption{Neutron stars with maximum masses in GR. The density values correspond to the critical central density $\rho_{c} = \rho_{crit}$ for which the total gravitational mass $M$ as a function of $\rho_{c}$ is a maximum.
}
\label{table1}
\begin{tabular}[c]{l c c c}
     \hline
        EoS   &   $\rho_{c}$ [$10^{18}\ \text{kg}/ \text{m}^3$]   &   $R$  [\text{km}]   &   $M$ [$M_\odot$]\\
	\hline
  SLy  &  2.858  &  9.981  &  2.046  \\
  BSk19  &  3.477  &  9.103  &  1.859  \\
  BSk21  &  2.290  &  11.041  &  2.272  \\	
	\hline
	\end{tabular}
\end{table}

\begin{figure*}
 \includegraphics[width=8.7cm]{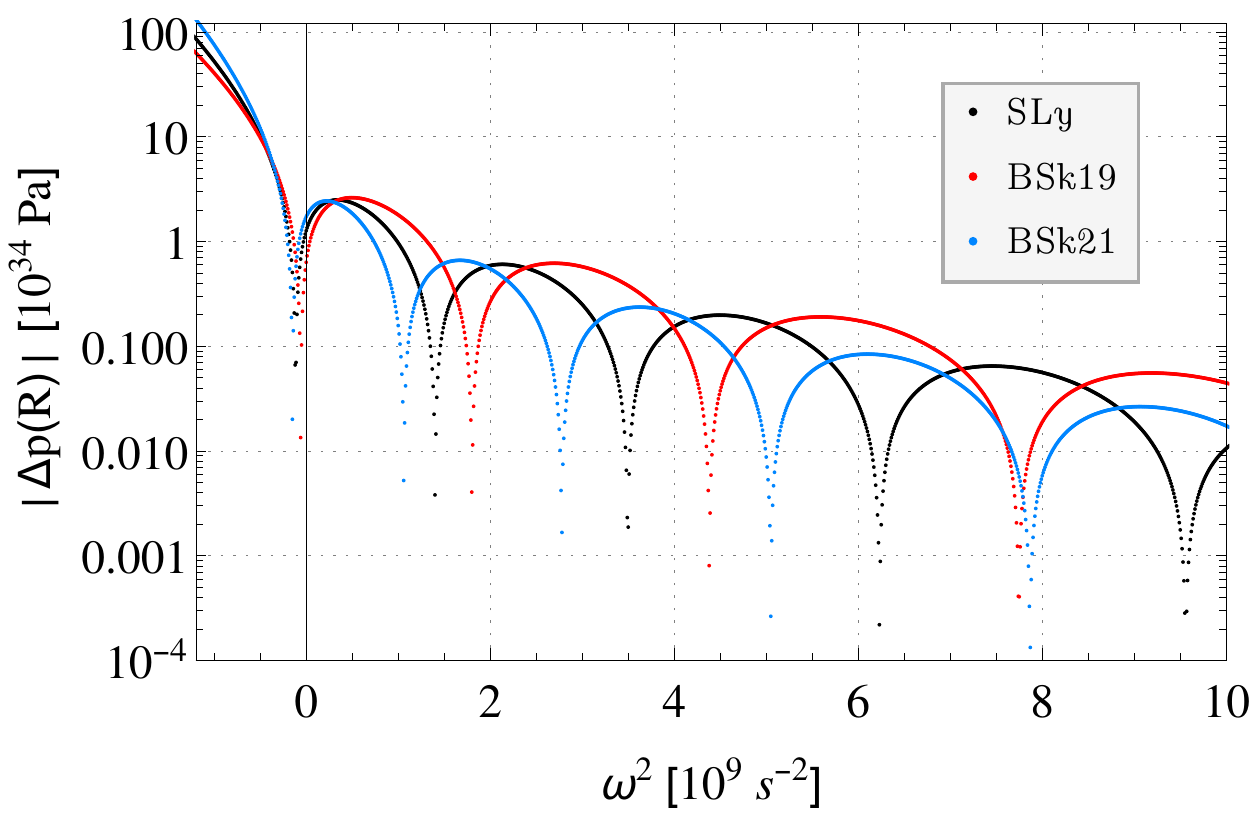} \
 \includegraphics[width=8.5cm]{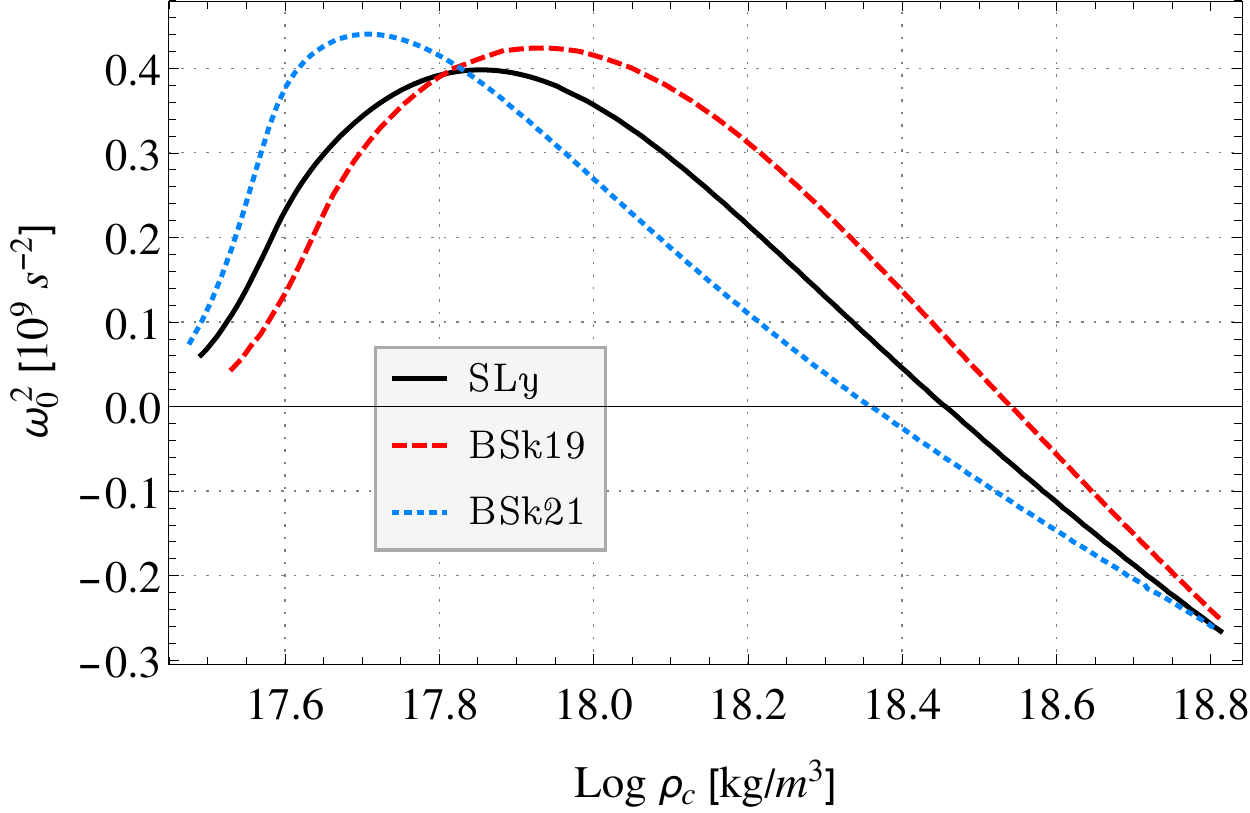}
 \caption{Left panel: Absolute value of the Lagrangian perturbation of the pressure at the surface on a logarithmic scale for a set of trial values of $\omega^2$ with a central mass density $\rho_c = 4.0 \times 10^{18}\ \text{kg}/\text{m}^3$. The minima in each curve correspond to the correct frequencies of the oscillation modes for equilibrium configurations. Clearly the three neutron stars are unstable. Right panel: Squared frequency of the fundamental mode as a function of the central mass density. According to the right panel of figure \ref{figure2}, the two methods for stellar stability are in agreement by indicating for each EoS from what critical density the stars are unstable.}  
 \label{figure3} 
\end{figure*}

The numerical integration of differential equations (\ref{13}) and (\ref{14}) is carried out using the shooting method, that is, we integrate the equations for a set of trial values of $\omega^2$ satisfying the condition (\ref{15}). In addition, we consider that normalized eigenfunctions correspond to $\zeta(0) = 1$ at the origin, and we integrate to the stellar surface. The values of the frequency $\omega$ for which the boundary condition (\ref{16}) is satisfied are the correct frequencies of the radial pulsations. In particular, for a central mass density $\rho_c = 4.0 \times 10^{18}\ \text{kg}/\text{m}^3$, we show in the left panel of Fig. \ref{figure3} the Lagrangian perturbation of the pressure for a set of test values $\omega^2$, where each minimum indicates the appropriate frequency. As a consequence, for a given star there are different eigenvalues $\omega_0^2 < \omega_1^2 < \cdots <\omega_n^2 < \cdots$ with their respective eigenfunctions $\zeta_n(r)$ and $\Delta p_n(r)$, where $n$ represents the number of nodes inside the star. The eigenvalue corresponding to $n=0$ is the fundamental mode, has the shortest frequency, and has no nodes between the center and the surface, whereas the first overtone $(n=1)$ has a node, the second overtone $(n=2)$ has two, and so on. We calculate the frequencies $f_n = \omega_n/2\pi$ of the first two radial modes for some values of central mass density, which are presented in tables \ref{table2}, \ref{table3}, and \ref{table4} for SLy, BSk19, and BSk21 equations of state, respectively. As in the Newtonian case, if $\omega^2 >0$, then $\omega$ is real and the eigenfunction $\varsigma(r)e^{i\omega t}$ is purely oscillatory (this is, the state of equilibrium is stable). On the other hand, for $\omega^2 <0$, the frequency is imaginary and the perturbations grow or decrease exponentially with time. This mean that for negative values of $\omega^2$, we have unstable radial oscillations \citep{BardeenTM, KokkotasRuoff}. Notice that if the fundamental mode of a star is stable ($\omega_0^2 > 0$), then all radial modes are stable. Indeed, according to the right panel of Fig. \ref{figure3}, the sufficient condition (i.e., $\omega^2 >0$) is also satisfied, yielding a good agreement between the two methods for stellar stability. This indicates that the maximum mass point and $\omega_0^2 = 0$ are found at the same value of central mass density.

\begin{figure*}
 \includegraphics[width=8.7cm]{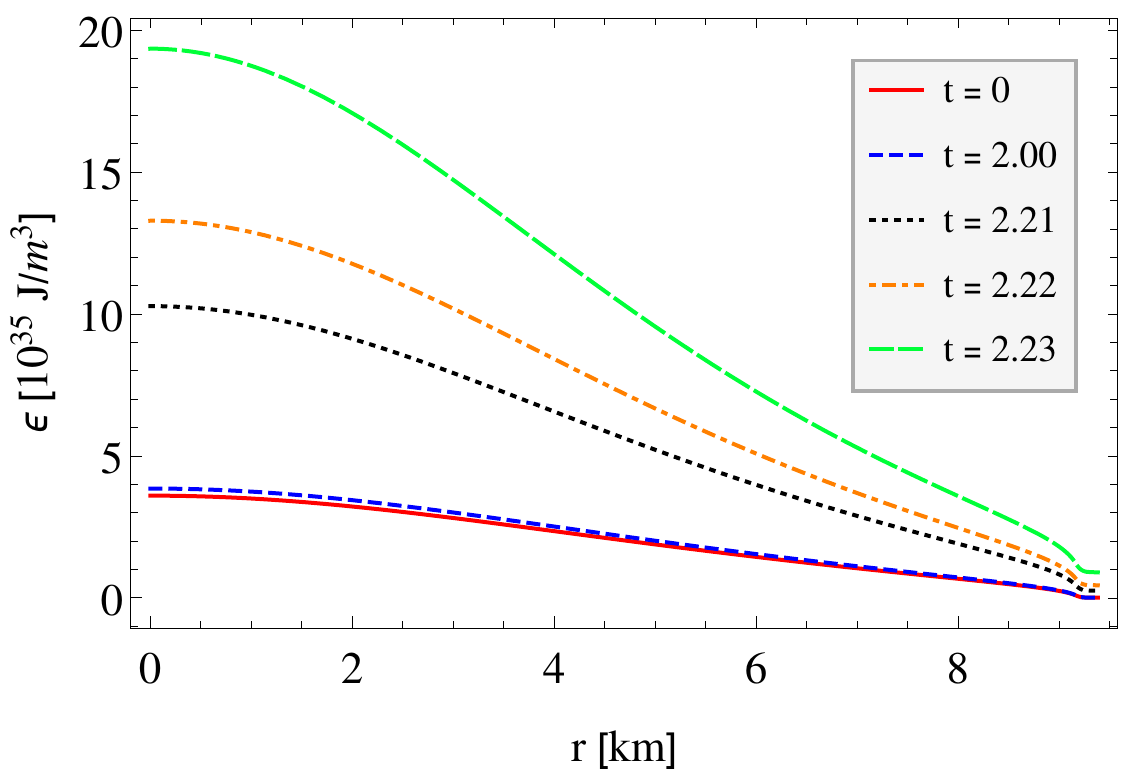} \
 \includegraphics[width=8.7cm]{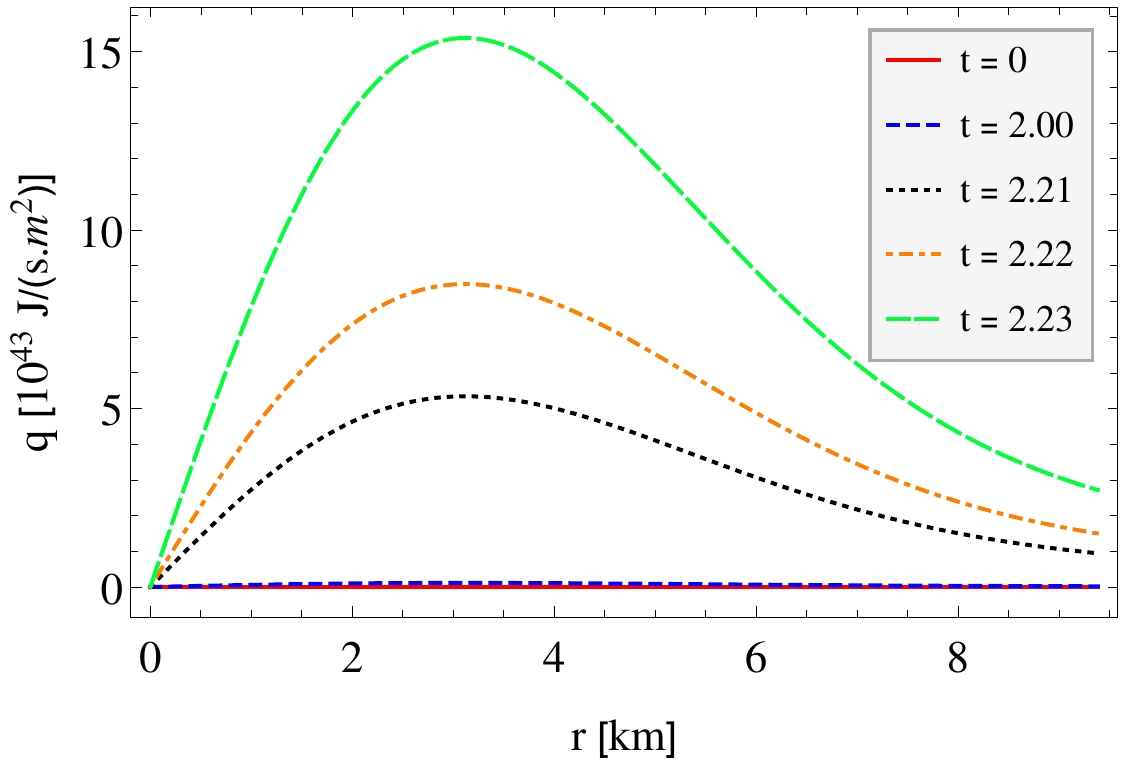}
 \caption{Energy density (left panel) and heat flux (right panel) as a function of the radial coordinate at different times (in $m$s units), for a central mass density $\rho_c = 4.0 \times 10^{18}\ \text{kg}/\text{m}^3$ with SLy EoS. This configuration corresponds to an unstable neutron star with radius $R = 9.384\ \text{km}$, initial mass $M = 2.015\ M_\odot$ and $m_{bh} = 1.278\ M_\odot$ at the end of the collapse.}
 \label{figure4} 
\end{figure*}

Stable neutron stars oscillate with a purely real (fundamental) frequency when are subjected to a radial compression or decompression, while the unstable stars (with central densities greater than critical density) suffer a gravitational collapse from rest to a black hole. In order to study the dynamical evolution of this implosion, we assume that the unstable configurations are initially in a state of hydrostatic equilibrium and then gradually begin to collapse until the formation of an event horizon. In the case of unstable neutron stars with SLy EoS, for an initial central mass density $\rho_c = 4.0 \times 10^{18}\ \text{kg}/\text{m}^3$, we solve equation (\ref{GC16}) with final conditions $f = 0.162$ and $df/dt = -9.402 \times 10^3\ \text{s}^{-1}$ at time $t =t_{bh} = -1.685 \times 10^{-5}\ \text{s}$. Then we can do a time displacement so that this configuration evolves from the instant $t = 0$ (when the interior structure is governed by TOV equations) until the moment of horizon formation $t_{bh} = 2.244\ m\text{s}$ (when the star has collapsed and the resulting black hole has a mass $m_{bh} = 1.278\ M_\odot$). The radial behavior of energy density (\ref{GC13}) and heat flux (\ref{GC15}) at different times are shown in the left and right panels of Fig. \ref{figure4}, respectively. The energy density always presents its maximum value at the stellar center and it changes considerably in the last instants of the collapse, while on the surface the change is relatively small. The radial heat flux undergoes great alterations in the intermediate zones of the collapsing star and its value is not zero on the surface. Notwithstanding that in the first two milliseconds these quantities remain almost constants, in the last 0.24 milliseconds there are relevant changes. This indicates that dissipative flows play an important role in the last stages of gravitational collapse.

\begin{figure*}
 \includegraphics[width=8.7cm]{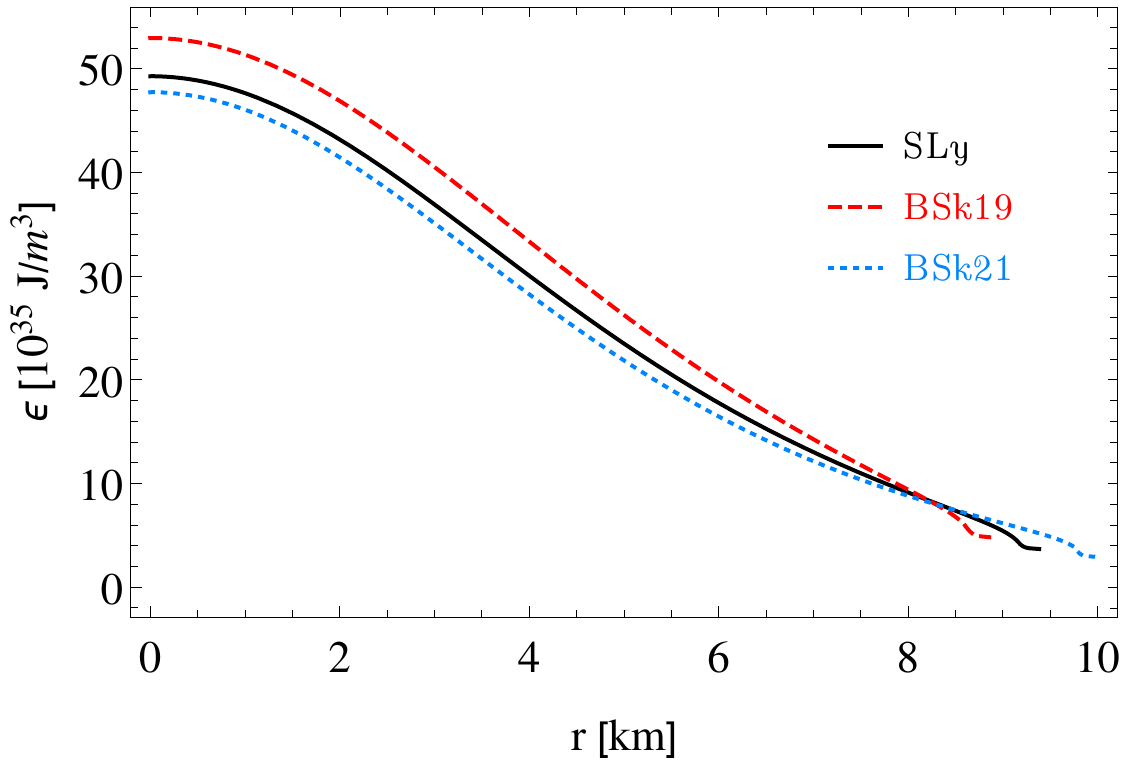} \ 
 \includegraphics[width=8.7cm]{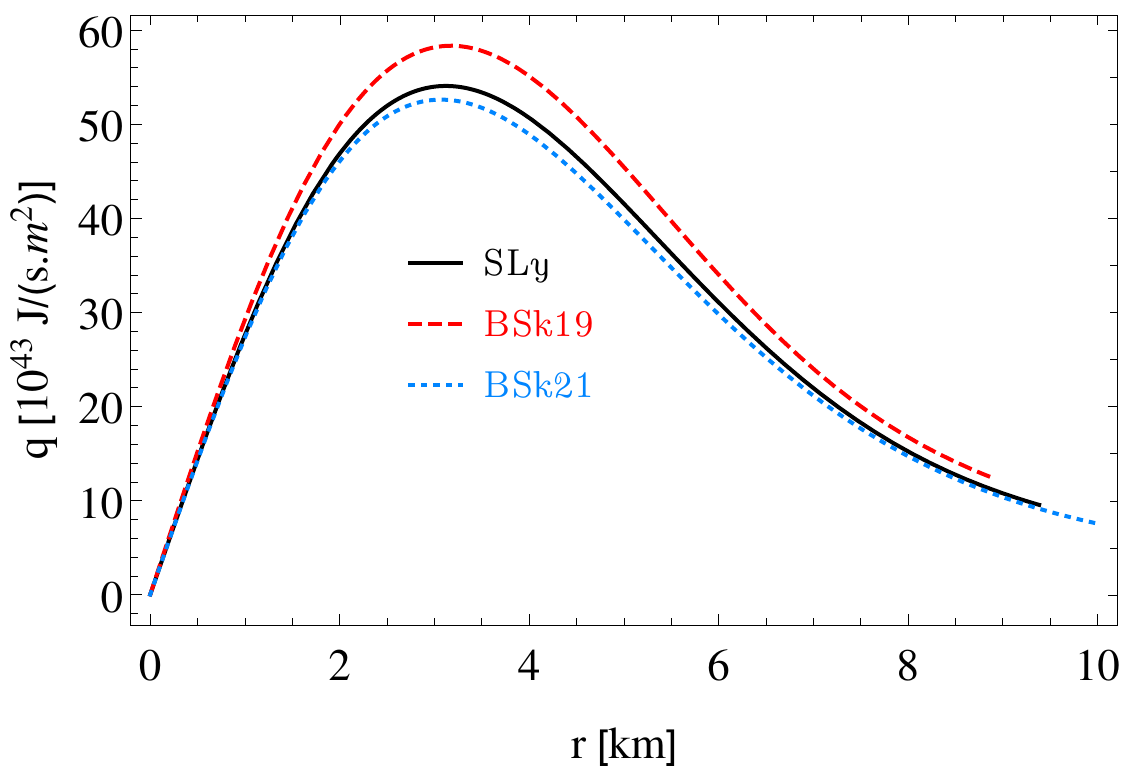}
 \caption{Radial profile of the energy density (left panel) and heat flux (right panel) at the time of horizon formation. The curves correspond to unstable neutron stars with initial central mass density $\rho_c = 4.0 \times 10^{18}\ \text{kg}/\text{m}^3$ for each EoS.}  
 \label{figure5} 
\end{figure*}

\begin{figure}
 \includegraphics[width=8.0cm]{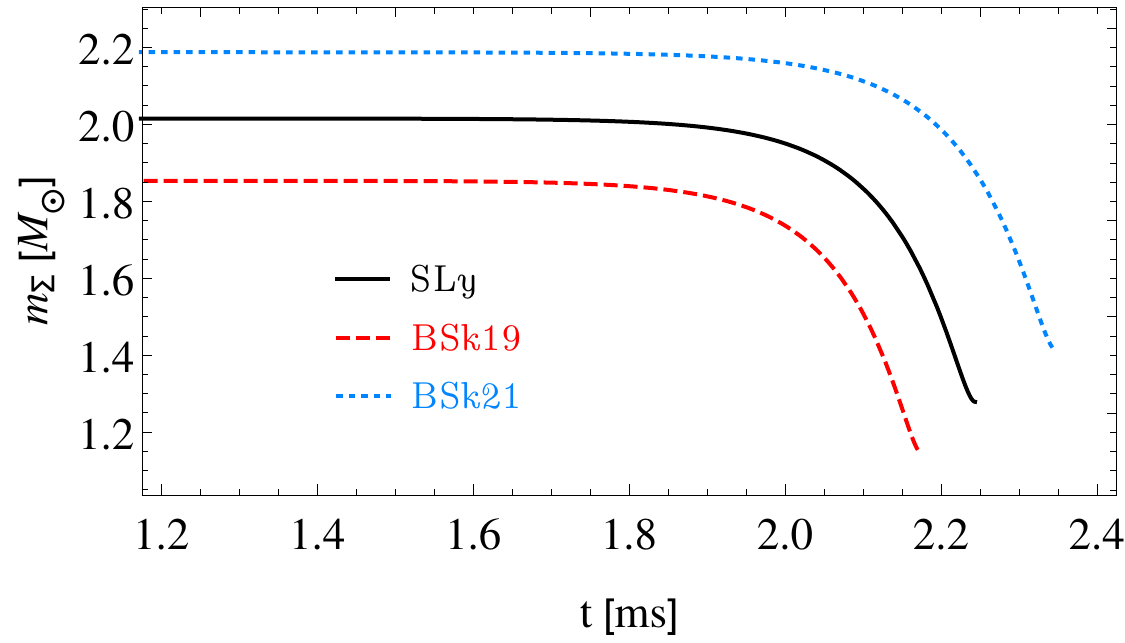}
 \includegraphics[width=8.0cm]{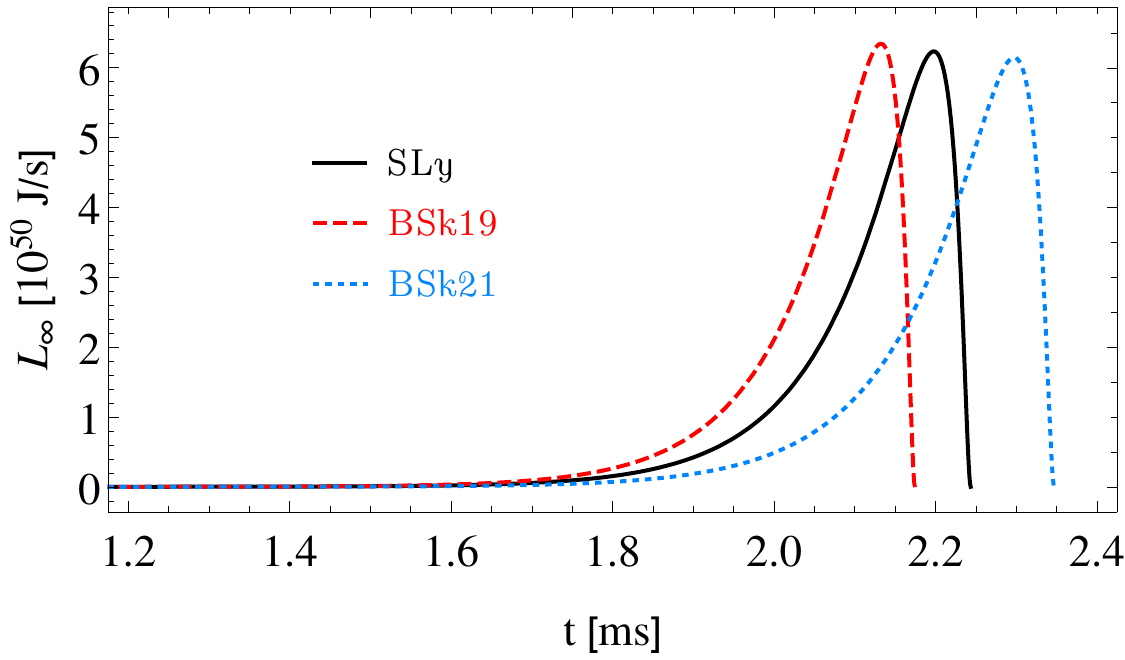}
 \includegraphics[width=8.0cm]{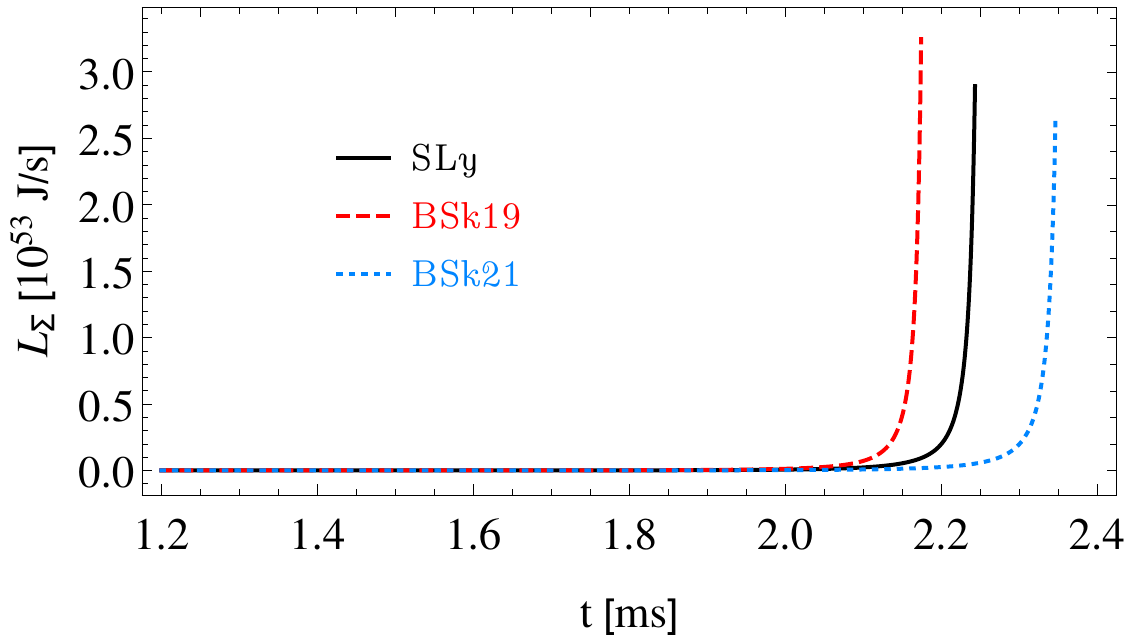}
 \caption{Temporal evolution of the total gravitational mass (upper panel), luminosity perceived by an observer at rest at infinity (intermediate panel), and luminosity as measured on its surface (lower panel) during the collapse process. For each unstable star the initial central mass density is the same as in figure \ref{figure5}.}
 \label{figure6} 
\end{figure}

Each EoS provides unstable stars with different parameters (in particular, different values of $f_{bh}$), so considering the same central density for the BSk19 and BSk21 equations of state, the radial profiles of the energy density and heat flux at time $t = t_{bh}$ are displayed in figure \ref{figure5}.  Moreover, in figure \ref{figure6} we illustrate the temporal evolution of the physical quantities given by (\ref{GC10b}), (\ref{GC11}), and (\ref{GC12}). The total gravitational mass (upper panel) decreases as the star collapses. This can be attributed to the fact that the star is emitting a large number of particles into outer spacetime. In table \ref{table5} we present the percentage of mass loss for some unstable configurations, revealing that at higher initial central density we have less mass loss regardless of the EoS. The luminosity perceived by an observer at rest at infinity (intermediate panel) has a maximum growth followed by an abrupt fall until the black hole formation. Ultimately, according to the lower panel of figure \ref{figure6} the luminosity measured on the stellar surface grows steeply in the final moments due to the fact that the gravitational redshift tends to infinity as the collapse approaches the horizon formation.

\begin{figure*}
 \includegraphics[width=8.4cm]{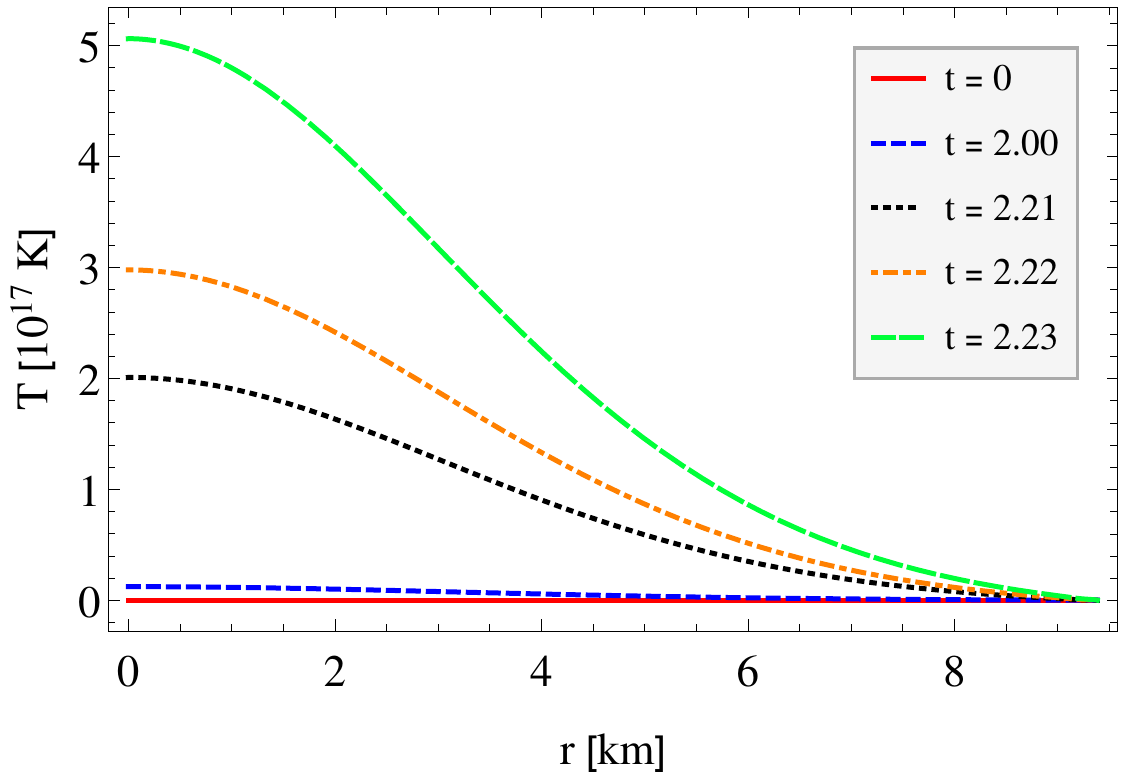} \ 
 \includegraphics[width=8.98cm]{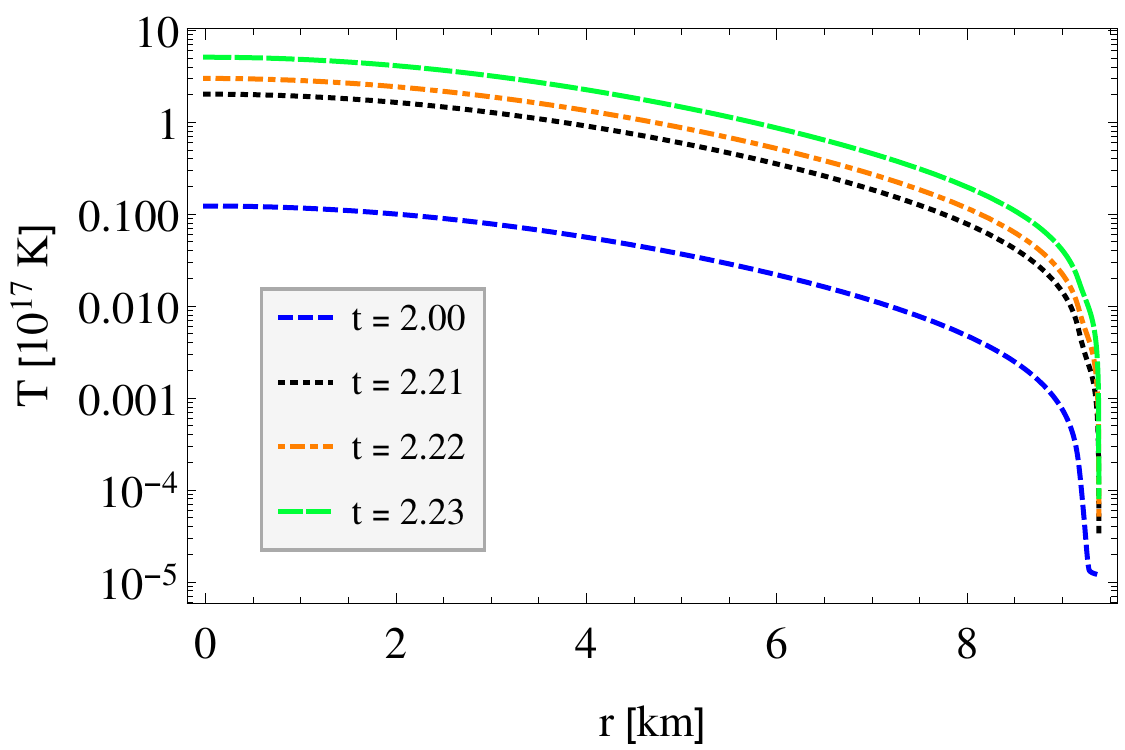}
 \caption{Left panel: Radial profile of the temperature at different times (in $m$s units) during the gravitational collapse. Right panel: The same radial behavior on a logarithmic scale where it can be seen that near the surface the temperature undergoes considerable changes. As we move away from the surface towards the stellar center, the temperature is $10^{5}$ times larger. We have used $\beta =5/2$ and $\alpha$ given by (\ref{GC33}), for an initial central mass density $\rho_c = 4.0 \times 10^{18}\ \text{kg}/\text{m}^3$ with SLy EoS.}  
 \label{figure7} 
\end{figure*}

\begin{figure*}
 \includegraphics[width=8.6cm]{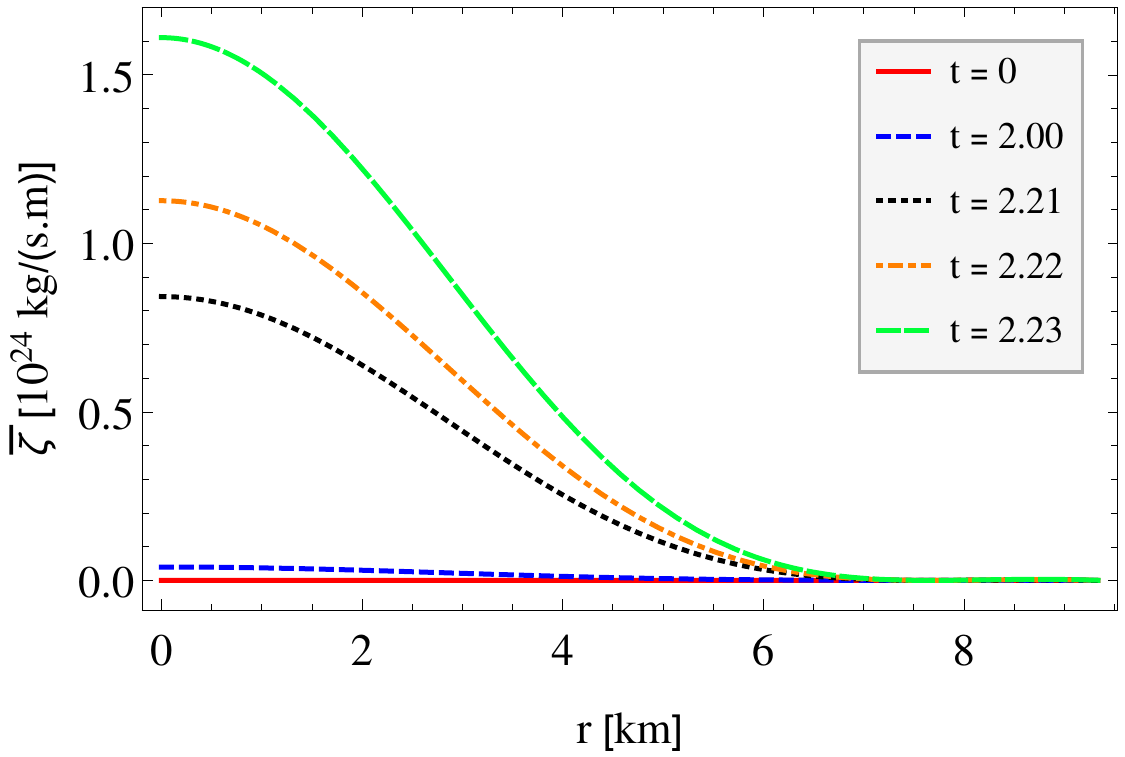} \ 
 \includegraphics[width=8.6cm]{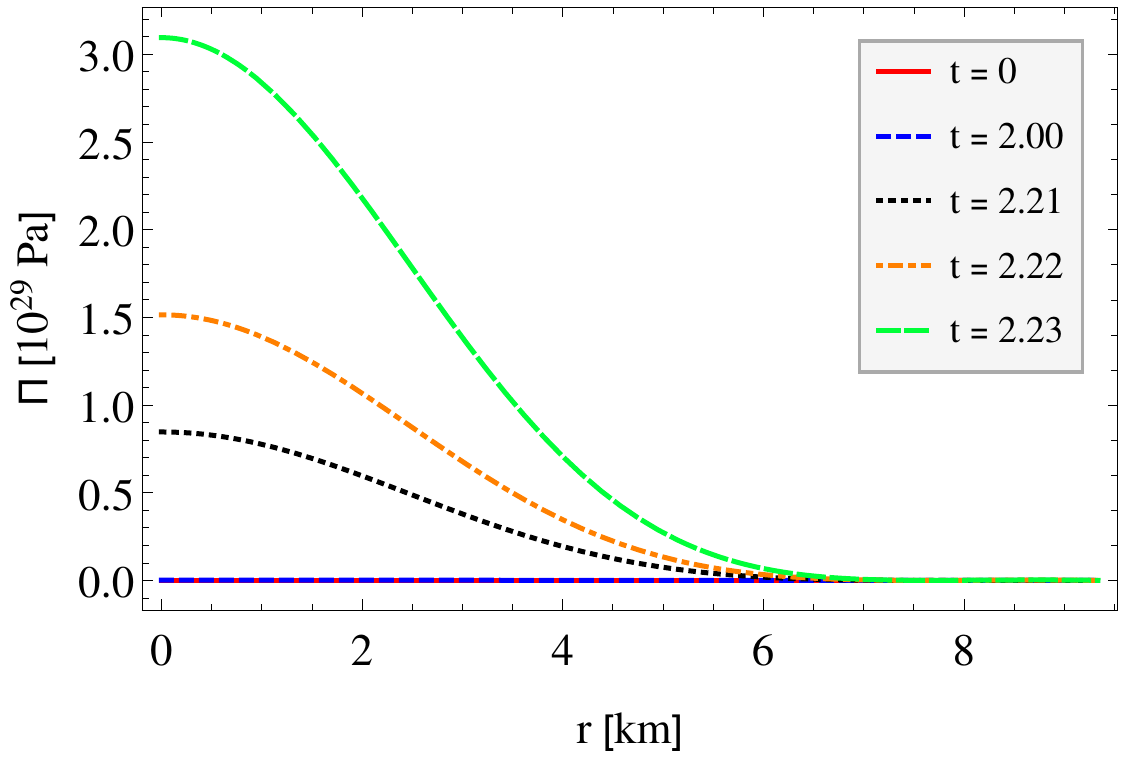}
 \caption{Bulk viscous coefficient (left panel) described by equation (\ref{GC30}) and bulk viscous pressure (right panel) given by (\ref{GC5a}) as functions of the radial coordinate at different times. The central density and parameters $\alpha$ and $\beta$ have the same values as in figure \ref{figure7}.}  
 \label{figure8} 
\end{figure*}

The thermal evolution of the gravitational collapse is obtained from equation (\ref{GC26}). For $\beta = 5/2$ and $\alpha$ given by the expression (\ref{GC33}), we can obtain a radial profile of the temperature at different times, which is plotted on both panels of  Fig. \ref{figure7}. It always presents its maximum value at the center, and as expected, it decreases as we approach the surface of the collapsing star. Furthermore, according to the right panel of the same figure, on a logarithmic scale we can see that near the stellar surface the temperature changes considerably since on the surface it is of the order of $10^{12}\ \rm K$. These sudden changes are associated with the behavior of the energy density near the surface through equation (\ref{GC33}). In the left panel of Fig. \ref{figure8} we display the bulk viscous  coefficient, indicating that it assumes its maximum value at the origin, but at about $2.5\ \rm km$ before reaching the surface it vanishes for any instant of time. The bulk viscous pressure defined by (\ref{GC5a}) is plotted in the right panel of the same figure, and we notice that it has a behavior similar to the bulk viscous coefficient. In the static limit the pressure is of the order of $10^{35}\ \rm Pa$, so that we can already perceive that the bulk viscous pressure is very small compared to the first term on the right-hand side of equation (\ref{GC14}).

\begin{figure*}
 \includegraphics[width=8.62cm]{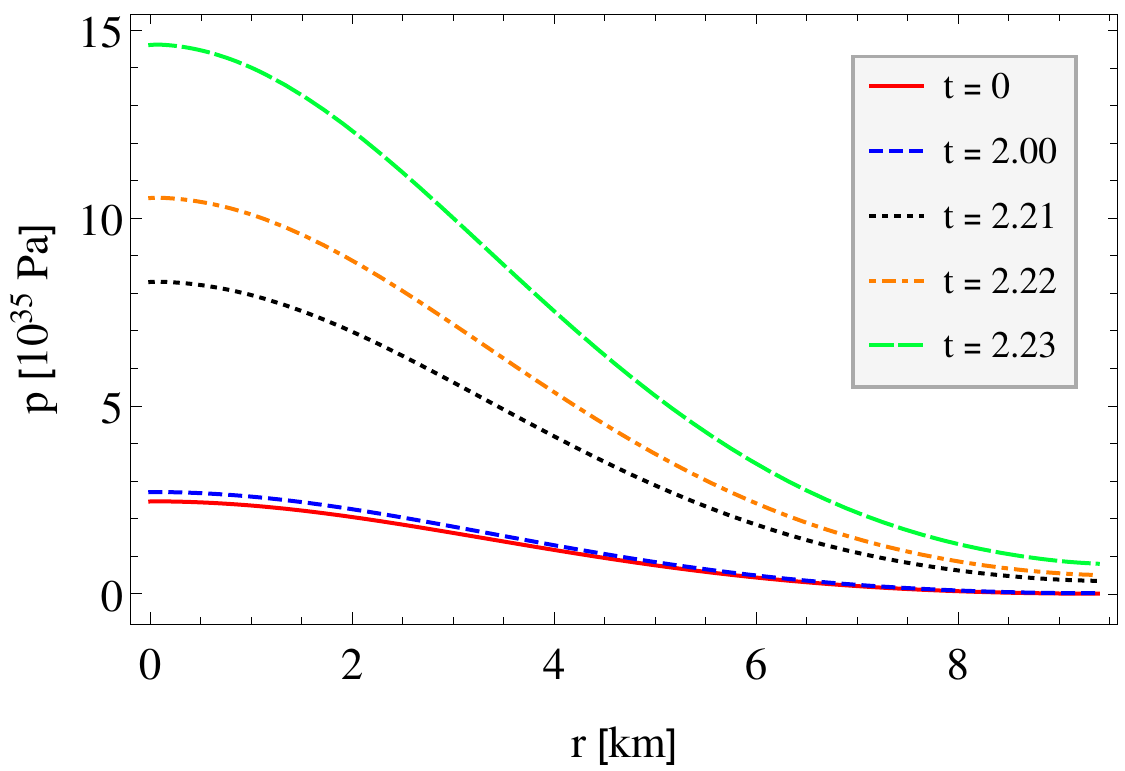} \ 
 \includegraphics[width=8.6cm]{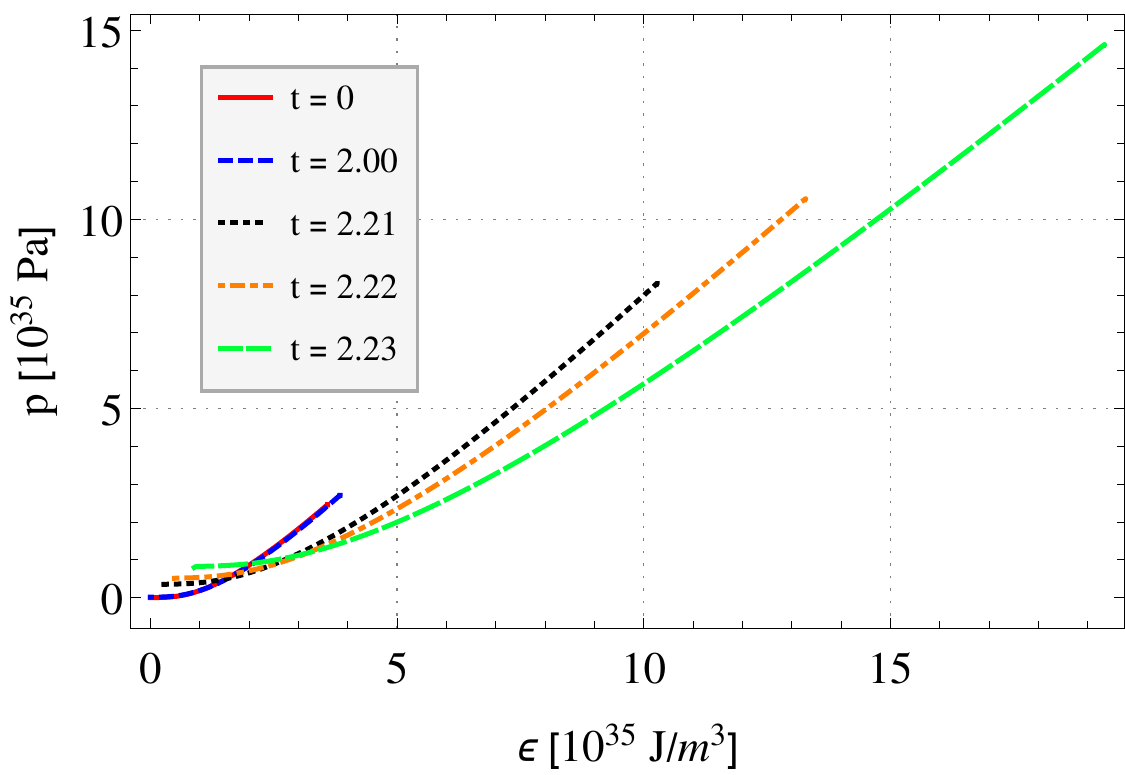}
 \caption{ Left panel: Radial profile of the pressure given by (\ref{GC14}). Right panel: Evolution of the SLy equation of state during the gravitational collapse by considering the same parameters used in figure \ref{figure7}.}  
 \label{figure9} 
\end{figure*}

\begin{figure}
 \includegraphics[width=7.2cm]{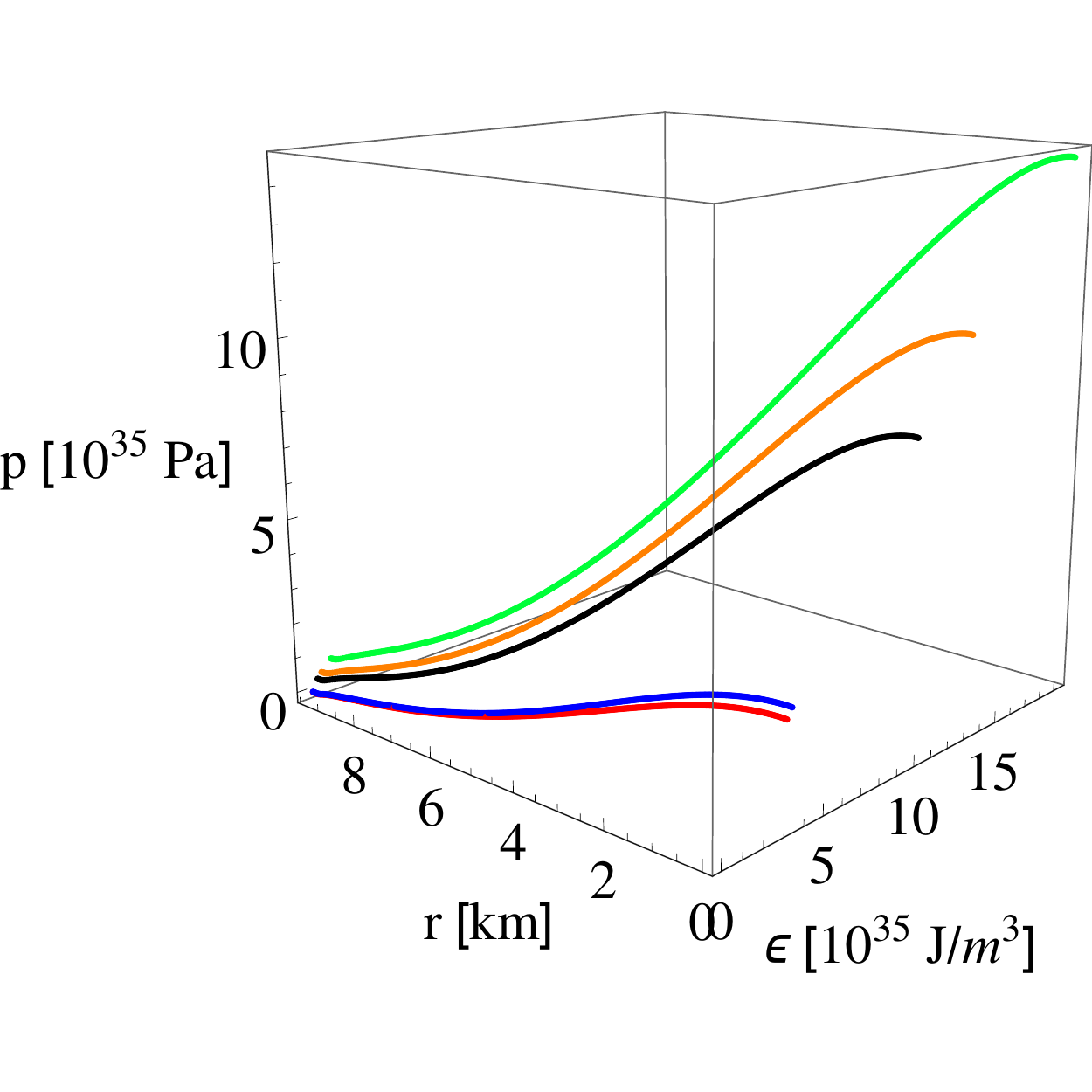}
 \caption{The same plot shown in the right panel of figure \ref{figure9}, but now including the radial coordinate. }  
 \label{figure10} 
\end{figure}

\begin{figure*}
 \includegraphics[width=6.4cm]{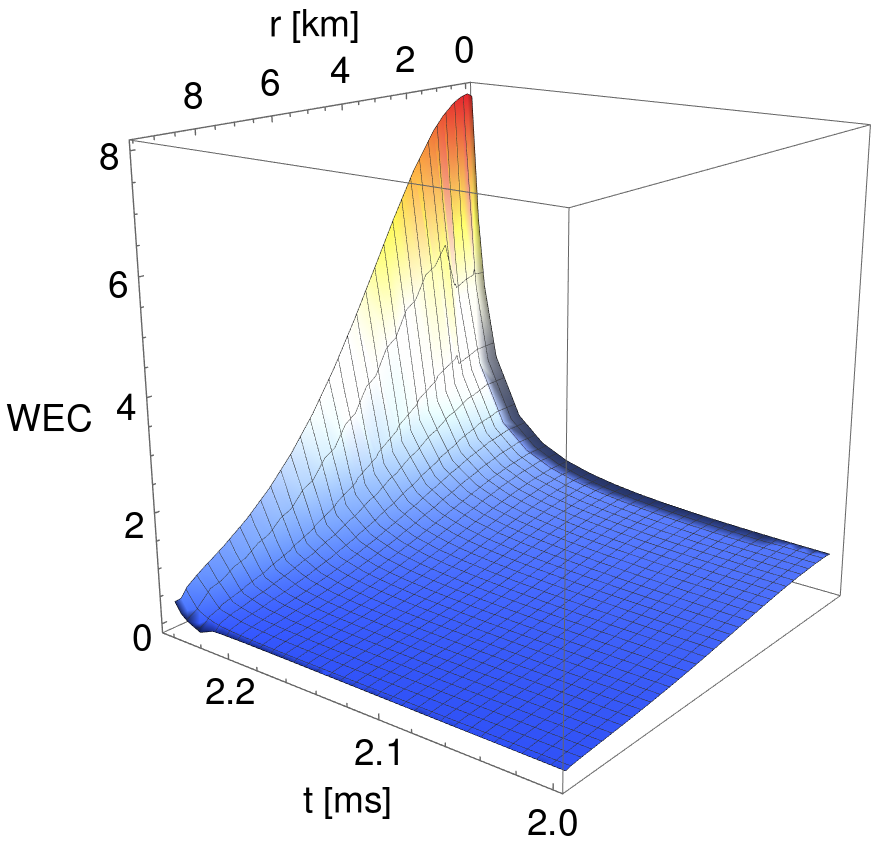} \ 
 \includegraphics[width=7.4cm]{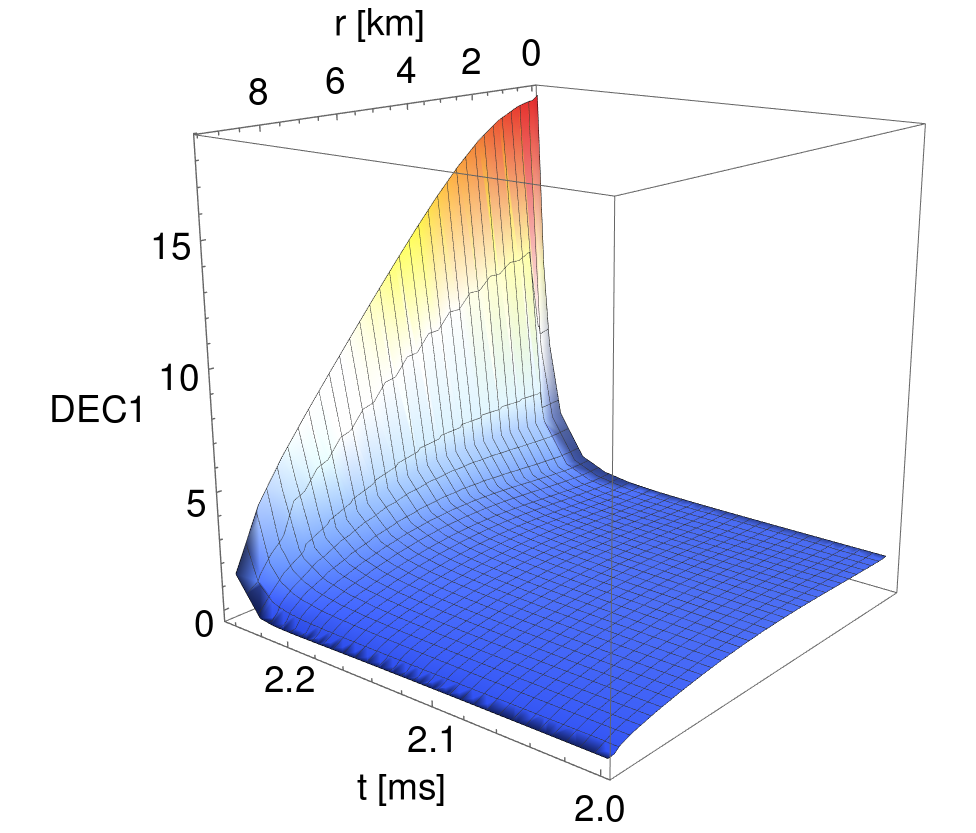} 
 \includegraphics[width=6.6cm]{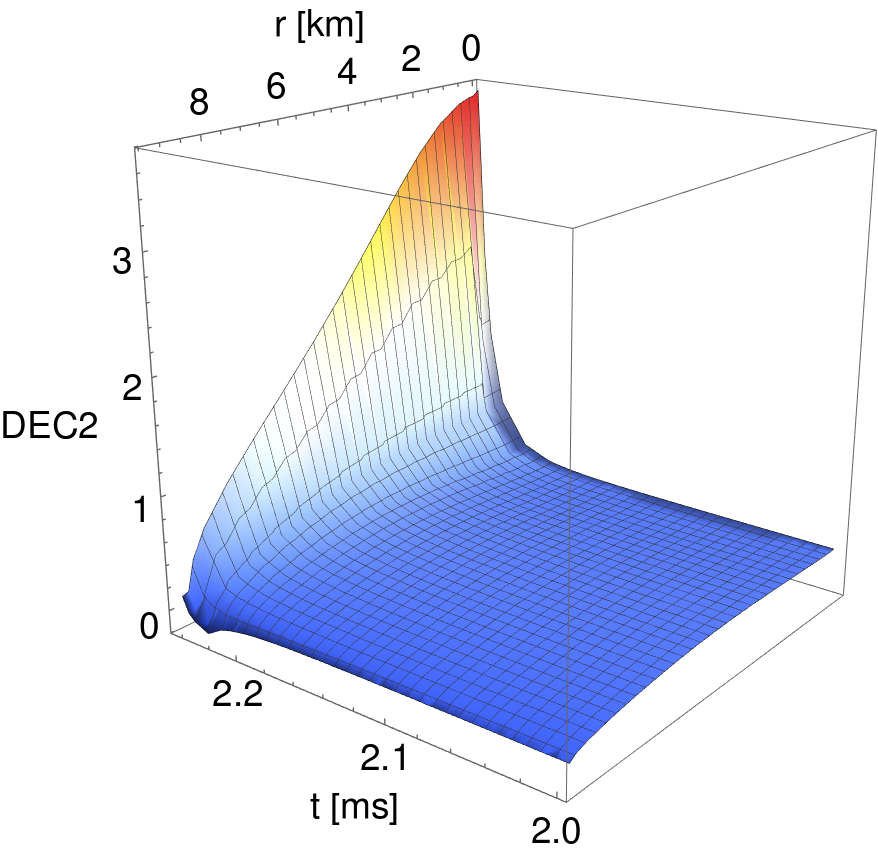} \ 
 \includegraphics[width=7.4cm]{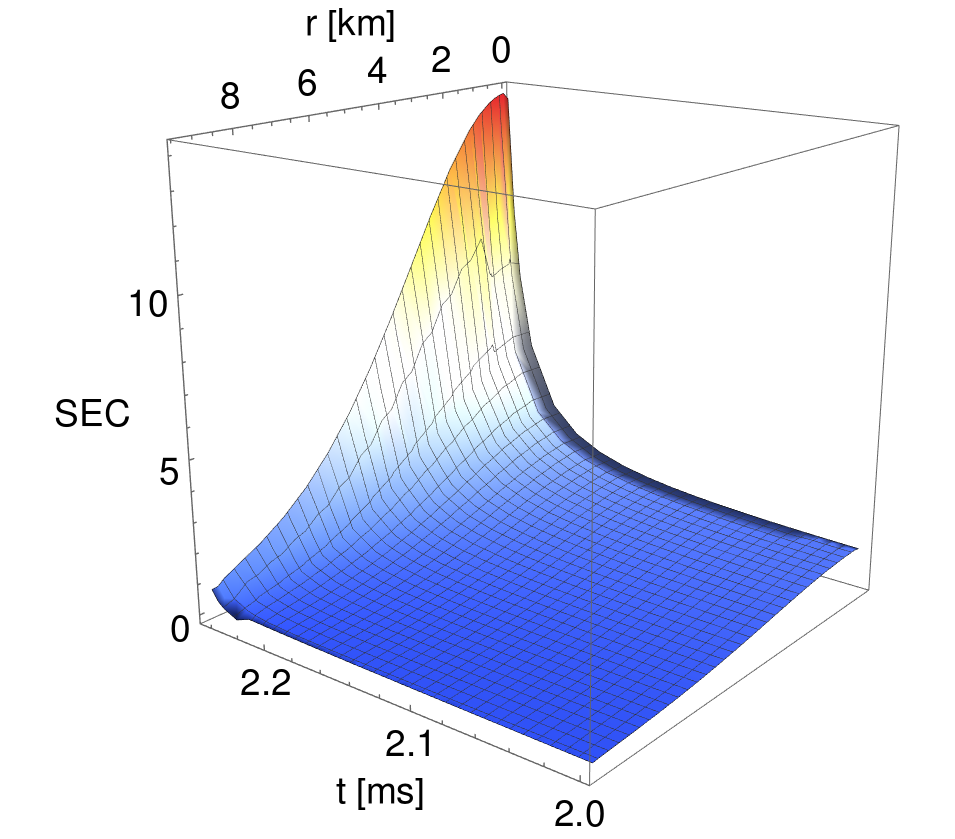}
 \caption{Energy conditions (\ref{GC36a}), (\ref{GC37a}), (\ref{GC37b}) and (\ref{GC38}), for a initial central mass density $\rho_c = 4.0 \times 10^{18}\ \text{kg}/\text{m}^3$ with SLy EoS.} \label{figure11} 
\end{figure*}

The radial profile of the pressure at different times is illustrated in the left panel of Fig. \ref{figure9}. At every instant of time the pressure always increases as we approach the horizon formation and decreases from the center to the surface. One can see that the pressure at the surface is no longer zero as in the static case, and this is because there exist a contribution from the radial heat flux and from the bulk viscous pressure, as indicated by equation (\ref{GC10c}). Gathering the results obtained for the energy density and pressure, we can investigate how the EoS behaves as the star collapses. The right panel of Fig. \ref{figure9} indicates that the equation of state undergoes significant changes during the collapse of an unstable neutron star. The maximum and minimum values in each curve correspond to the center and the surface of the star, respectively. This can be better visualized in a three-dimensional graph as is presented in figure \ref{figure10}. Finally, the collapsing neutron star with initial central mass density $\rho_c = 4.0 \times 10^{18}\ \text{kg}/\text{m}^3$ and SLy EoS, is physically reasonable because it satisfies the energy conditions in the full extent of the star and throughout the collapse process. Indeed, the energy conditions (\ref{GC36a}), (\ref{GC37a}), (\ref{GC37b}) and (\ref{GC38}), denoted by:
\begin{align*}
    \text{WEC} &= \Delta \ \ [10^{36}\ \rm J \cdot m^{-3}] ,  \\
    \text{DEC1} &= \epsilon - p - \Pi \ \ [10^{35}\ \rm J \cdot m^{-3}] ,  \\
    \text{DEC2} &= \epsilon + \Delta - 3(p + \Pi) \ \ [10^{36}\ \rm J \cdot m^{-3}] ,  \\
    \text{SEC} &= \epsilon + 2(p+ \Pi) \ \ [10^{36}\ \rm J \cdot m^{-3}] ,
\end{align*}
are presented in Fig. \ref{figure11}, respectively. It is important to mention that we have tested this procedure for the other values of initial central density shown in table \ref{table5}, obtaining a similar behavior during the evolution of the collapse. 

\begin{table}
\centering
\caption{Percentage of mass loss during the collapse process of unstable stars for each EoS, this is $\frac{(M - m_{bh})}{M}\times 100 \%$.
}
\label{table5}
\begin{tabular}[c]{c c c c}
     \hline
     $\rho_{c}$ [$10^{18}\ \text{kg}/ \text{m}^3$]   &   SLy   &   BSk19   &   BSk21  \\
	\hline
  2.50  &  --  &  --  &  38.25   \\
  3.00  &  38.91  &  --  &  36.79   \\
  3.50  &  37.46  &  39.59  &  35.89   \\
  4.00  &  36.58  &  38.15  &  35.39   \\
  5.00  &  35.52  &  36.47  &  34.84   \\
  6.00  &  35.10  &  35.64  &  34.70   \\	
	\hline
	\end{tabular}
\end{table}

\section{Conclusions}\label{sec:level5} 

Within the framework of general relativity, in this work we have studied the stability against radial oscillations and the dynamical gravitational collapse of neutrons stars for three realistic equations of state. To check if the equilibrium configurations are stable or unstable with respect to a radial perturbation, we solved the equations that govern the radial pulsations and calculated the frequencies of the vibration modes. Then we proceed to study the temporal evolution of the gravitational collapse of unstable stars to the moment of horizon formation. This was achieved by introducing a time dependence on metric functions so that under a certain limit we could recover the static case. The junction conditions and dynamical equations that describe some relevant physical quantities of neutron stars undergoing dissipative gravitational collapse were derived.  

The radial heat flow plays a fundamental role during the collapse, it allowed us to obtain a temperature profile using the relativistic thermodynamic formalism developed by Eckart for heat transport. Consequently, we have investigated the temporal and radial behavior of the bulk viscous coefficient as well as the bulk viscous pressure. Once the energy density and pressure were known, we were able to examine how the equation of state evolves as an unstable neutron star collapses. Finally, we emphasize that our stellar collapse model satisfies all energy conditions for all extent of the star and throughout the collapse process, which is crucial for the physical validity of the system.

\section*{Acknowledgements}

JMZP thanks Brazilian funding agency CAPES for PhD scholarship 331080/2019. The author (MFAdaS) acknowledges the financial support from Financiadora de Estudos e  Projetos - FINEP - Brazil, Funda{\c c}{\~a}o de Amparo {\`a} Pesquisa do Estado do Rio de Janeiro - FAPERJ - Brazil and Conselho Nacional de Desenvolvimento Cient{\'i}fico e Tecnol{\'o}gico - CNPq - Brazil.




\bibliographystyle{mnras}
\bibliography{example} 




\appendix

\appendix

\section{Parameters and oscillation spectrum of neutron stars with (\ref{10}) and (\ref{11}) EoS}

In this appendix we provide three tables of numerical data corresponding to each equation of state. For some values of central mass density, we present the radius, total mass, frequency of the fundamental mode and the first overtone, as well as the mass of the black hole formed for the unstable configurations. In the case of unstable stars, when $\omega_0^2$ is negative, the frequency is imaginary and we are denoting it by an asterisk.

\begin{table}
\centering
\caption{Data for the SLy EoS.
}
\label{table2}
\begin{tabular}[c]{cccccc}
     \hline
$\rho_{c}$    &    $R$    &    $M$   &   $f_0$   &    $f_1$   &   $m_{bh}$  \\
$[10^{18} \text{kg}/ \text{m}^3]$   &   [\text{km}]   &   [$M_\odot$]   &   [kHz]   &   [kHz]   &   [$M_\odot$]   \\
	\hline
  0.500  &  12.116  &  0.535  &  2.945  &  4.179  &  --  \\
  0.800  &  11.817  &  1.118  &  3.156  &  6.707  &  --  \\
  1.000  &  11.689  &  1.417  &  3.009  &  7.011  &  --  \\
  1.500  &  11.183  &  1.836  &  2.440  &  6.984  &  --  \\
  2.000  &  10.666  &  1.991  &  1.814  &  6.750  &  --  \\
  2.500  &  10.236  &  2.040  &  1.102  &  6.515  &  --  \\
  2.800  &  10.020  &  2.046  &  0.427  &  6.385  &  --  \\
  3.000  &  9.890  &  2.046  &  0.660*  &  6.305  &  1.250  \\
  3.500  &  9.612  &  2.034  &  1.336*  &  6.121  &  1.272  \\
  4.000  &  9.384  &  2.015  &  1.708*  &  5.959  &  1.278  \\
  5.000  &  9.038  &  1.971  &  2.176*  &  5.686  &  1.271  \\
  6.000  &  8.788  &  1.929  &  2.480*  &  5.462  &  1.252  \\
	\hline
	\end{tabular}
\end{table}

\begin{table}
\centering
\caption{Data for the BSk19 EoS.
}
\label{table3}
\begin{tabular}[c]{cccccc}
     \hline
$\rho_{c}$    &    $R$    &    $M$   &   $f_0$   &    $f_1$   &   $m_{bh}$  \\
$[10^{18} \text{kg}/ \text{m}^3]$   &   [\text{km}]   &   [$M_\odot$]   &   [kHz]   &   [kHz]   &   [$M_\odot$]   \\
	\hline
  0.500  &  11.874  &  0.388  &  2.768  &  3.343  &  --  \\
  0.800  &  11.136  &  0.832  &  3.276  &  6.428  &  --  \\
  1.000  &  10.995  &  1.090  &  3.247  &  7.116  &  --  \\
  1.500  &  10.576  &  1.520  &  2.892  &  7.494  &  --  \\
  2.000  &  10.120  &  1.726  &  2.410  &  7.420  &  --  \\
  2.500  &  9.713  &  1.816  &  1.878  &  7.257  &  --  \\
  3.000  &  9.372  &  1.851  &  1.257  &  7.082  &  --  \\
  3.200  &  9.253  &  1.856  &  0.942  &  7.013  &  --  \\
  3.500  &  9.091  &  1.859  &  0.270*  &  6.913  &  1.123  \\
  4.000  &  8.857  &  1.853  &  1.221*  &  6.757  &  1.146  \\
  5.000  &  8.495  &  1.826  &  1.950*  &  6.483  &  1.160  \\
  6.000  &  8.229  &  1.793  &  2.371*  &  6.252  &  1.154  \\
	\hline
	\end{tabular}
\end{table}

\begin{table}
\centering
\caption{Data for the BSk21 EoS.
}
\label{table4}
\begin{tabular}[c]{cccccc}
     \hline
$\rho_{c}$    &    $R$    &    $M$   &   $f_0$   &    $f_1$   &   $m_{bh}$  \\
$[10^{18} \text{kg}/ \text{m}^3]$   &   [\text{km}]   &   [$M_\odot$]   &   [kHz]   &   [kHz]   &   [$M_\odot$]   \\
	\hline
  0.500  &  12.361  &  0.766  &  3.339  &  5.345  &  --  \\
  0.800  &  12.589  &  1.546  &  2.969  &  6.960  &  --  \\
  1.000  &  12.458  &  1.852  &  2.616  &  6.811  &  --  \\
  1.500  &  11.863  &  2.183  &  1.808  &  6.345  &  --  \\
  2.000  &  11.307  &  2.265  &  0.995  &  5.988  &  --  \\
  2.200  &  11.117  &  2.272  &  0.531  &  5.872  &  --  \\
  2.500  &  10.864  &  2.269  &  0.797*  &  5.719  &  1.401  \\
  3.000  &  10.514  &  2.248  &  1.374*  &  5.508  &  1.421  \\
  3.500  &  10.233  &  2.218  &  1.705*  &  5.334  &  1.422  \\
  4.000  &  10.003  &  2.187  &  1.941*  &  5.187  &  1.413  \\
  5.000  &  9.652  &  2.127  &  2.273*  &  4.946  &  1.386  \\
  6.000  &  9.396  &  2.075  &  2.505*  &  4.752  &  1.355  \\
	\hline
	\end{tabular}
\end{table}


\bsp	
\label{lastpage}
\end{document}